\documentclass[reqno,a4paper]{amsart}
\usepackage[T1]{fontenc}
\usepackage{comment,amssymb,latexsym,upref,enumerate,fouridx}
\usepackage[normalem]{ulem}
\usepackage{dutchcal}
\usepackage{mathrsfs,xcolor}
\usepackage{graphicx}
\graphicspath{ {./images/} }
\usepackage{subfigure}
\usepackage{placeins}
\usepackage{centernot}
\usepackage[colorlinks,linkcolor=blue,citecolor=blue]{hyperref}

\allowdisplaybreaks
\numberwithin{equation}{section}

\theoremstyle{plain}
\newtheorem{thm}{Theorem}[section]

\theoremstyle{definition}

\theoremstyle{remark}
\newtheorem{rem}[thm]{Remark}

\newcommand*{\dd}{\mathrm{d}}

\input Tdef.def

\newcounter{mnotecount}[section]
\let\oldmarginpar\marginpar
\renewcommand\marginpar[1]{\-\oldmarginpar[\raggedleft\footnotesize #1]%
 {\raggedright\footnotesize #1}}

\graphicspath{{./}{./Images/Convergence Plots/}{./Images/Asymptotic Comparison/}{./Images/Density Contrast/}{./Images/Rho Plots/}{./Images/}}

\begin{document}
\title[Future instability of FLRW fluid solutions for linear equations of state $p=K\rho$ with $1/3 <K<1$]{Future instability of FLRW fluid solutions for linear equations of state $p=K\rho$ with $1/3 <K<1$} 

\author[F.~Beyer]{Florian Beyer}
\address{Dept of Mathematics and Statistics\\
730 Cumberland St\\
University of Otago, Dunedin 9016\\ New Zealand}
\email{fbeyer@maths.otago.ac.nz}

\author[E.~Marshall]{Elliot Marshall}
\address{School of Mathematics\\
9 Rainforest Walk\\
Monash University, VIC 3800\\ Australia}
\email{elliot.marshall@monash.edu}

\author[T.A.~Oliynyk]{Todd A.~Oliynyk}
\address{School of Mathematics\\
9 Rainforest Walk\\
Monash University, VIC 3800\\ Australia}
\email{todd.oliynyk@monash.edu}

\begin{abstract} 
Using numerical methods, we examine the dynamics of nonlinear perturbations in the expanding time direction, under a Gowdy symmetry assumption, of FLRW fluid solutions to the Einstein-Euler equations with a positive cosmological constant $\Lambda>0$ and a linear equation of state $p = K\rho$ for the parameter values $1/3<K<1$. This paper builds upon the numerical work in \cite{Marshalloliynyk:2022} in which the simpler case of a fluid on a fixed FLRW background spacetime was studied. The numerical results presented here confirm that the instabilities observed in \cite{Marshalloliynyk:2022} are also present when coupling to gravity is included as was previously conjectured in \cite{Rendall:2004,Speck:2013}. In particular, 
for the full parameter range $1/3 < K <1$, we find
that the fractional density gradient 
of the nonlinear perturbations develop steep gradients near a finite number of spatial points and becomes unbounded there
at future timelike infinity. 
\end{abstract}

\maketitle

\section{Introduction\label{intro}}
Beginning with the seminal work of Friedrich \cite{Friedrich:1986}, the future (i.e.~expanding) stability of cosmological solutions on exponentially expanding spacetimes has been the source of much research. Recently, on account of their
importance in modern standard cosmology \cite{mukhanov2005}, the future stability of fluid filled 
cosmologies with linear equations of
state, $p=K\rho$, have been intensively
studied with the first rigorous results
due to Rodnianski and Speck \cite{RodnianskiSpeck:2013,Speck:2012} who proved the future stability of nonlinear perturbations of FLRW (i.e.~spatially homogeneous and isotropic)  solutions to the Einstein-Euler equations with a positive cosmological constant for the parameter range $0<K<1/3$. Stability results for the end points $K=1/3$ and $K=0$ were subsequently established in \cite{LubbeKroon:2013} and \cite{HadzicSpeck:2015},
respectively. Related works have also examined different approaches to establishing stability \cite{Friedrich:2017,LiuOliynyk:2018b,LiuOliynyk:2018a,Oliynyk:CMP_2016}, fluids with nonlinear equations of state \cite{LeFlochWei:2021,LiuWei:2021}, and other expanding spacetimes (such as power-law expansion) \cite{FOW:2021,Ringstrom:2009,Speck:2013,Wei:2018}.

The question of stability for the parameter range $1/3<K<1$, until recently, remained an open question. In fact, it was widely expected that solutions to the Einstein-Euler equations were unstable when $K>1/3$. This was primarily a result of the influential work of Rendall \cite{Rendall:2004} who used formal expansions
about future timelike infinity to investigate the asymptotic behaviour of relativistic fluids on exponentially expanding FLRW spacetimes. In particular, Rendall found that if $1/3<K<1$ and the leading order term in the expansion of the fluid's spatial velocity about timelike infinity vanished at any spatial point, then the formal expansions would become inconsistent. He speculated this was due to inhomogeneous features, so-called spikes, developing in the fluid density which would cause the fractional density gradient to blow-up at future timelike infinity. Another argument supporting the instability of solutions for $1/3<K<1$ was given by Speck \cite[\S 1.2.3]{Speck:2013} who identified certain terms in the equations that might dynamically drive the instability. 
Instability to the future of solutions to the Einstein-Euler equations for $1/3<K<1$ has also been observed in spherical symmetry \cite{harada_stability_2001}.

More recently, the work of the third author \cite{Oliynyk:2021} established the existence of a class of non-isotropic spatially homogeneous solutions to the relativistic Euler equations on fixed exponentially expanding FLRW background spacetimes that are \textbf{(i)} stable to the future under small nonlinear perturbations for $1/3<K<1/2$ and for which \textbf{(ii)} the initial data of the perturbations could be chosen arbitrarily close to the initial data of a spatially homogeneous and isotropic solution. 
While the second point implies that the solutions from \cite{Oliynyk:2021} can be viewed as perturbations of spatially homogeneous and isotropic solutions with zero spatial velocity, it should be noted that the spatial velocity of the fluids in \cite{Oliynyk:2021} must be non-vanishing everywhere and hence do not constitute  a general class of perturbations of spatially homogeneous and isotropic solutions.

In the article \cite{Marshalloliynyk:2022}
by the last two authors, the stability result of \cite{Oliynyk:2021} was improved to cover the whole parameter range $1/3<K<1$. Additionally, a numerical investigation of the stability to the future of the class of spatially homogeneous and isotropic solutions to the Euler equations on fixed FLRW vacuum solutions with positive cosmological constant was carried out. Specifically, numerical solutions of the relativistic Euler equations
under a $\Tbb^{2}$-symmetry assumption were constructed globally to the future for a class of initial data that included perturbations of spatially homogeneous and isotropic initial data for which the spatial velocity of the fluid vanished at a finite number of points on the initial hypersurface. It is important to emphasize that the vanishing of the fluid's spatial velocity means that these solutions do not satisfy the conditions of the stability theorem from \cite{Marshalloliynyk:2022}. 

The main conclusions from the numerical study carried out in \cite{Marshalloliynyk:2022} can be summarised as follows:
\begin{enumerate}[(1)]
    \item For each $K\in(1/3,1)$ and each choice of initial data sufficiently close to spatially homogeneous and isotropic data, the numerical solutions of the relativistic Euler equations display ODE behaviour at late times and are remarkably well-approximated by an \textit{asymptotic system} that is constructed by discarding all spatial derivatives from a particular formulation of the relativistic Euler equations; see \cite[\S3.2.2]{Marshalloliynyk:2022} for details.
    \item  For each $K\in(1/3,1)$ and each choice of initial data that is sufficiently close to spatially homogeneous initial data and for which the spatial velocity of the fluid vanishes initially at a finite number of points, the fractional density gradient of the fluid develops steep gradients near a finite number of spatial points where it becomes unbounded at future timelike infinity; see \cite[\S3.2.3]{Marshalloliynyk:2022} for details.
\end{enumerate}

The aim of the current article is to extend the numerical study of the $\Tbb^2$-symmetric relativistic Euler equations from \cite{Marshalloliynyk:2022} to include coupling to Einstein gravity in the case $1/3<K<1$ and thereby to verify quantitatively the conjectures from \cite{Rendall:2004, Speck:2013} regarding unstable dynamics. In order to accomplish this, we numerically evolve the Einstein-Euler equations with spatial $\Tbb^3$-topology under a Gowdy symmetry assumption (see Section~\ref{sec:EinsteinEulerderivationGowdy}). The Gowdy spacetimes we consider in this article are especially well-suited to both analytical and numerical treatments (e.g.~ \cite{ames2017,amorim2009,berger1993,BeyerHennig:2012,beyer2010b,beyer2017,beyer2020b,chrusciel1990,isenberg1990,kichenassamy1998,LeFlochRendall:2011,rendall2000,ringstrom2009a}) due to the presence of two Killing fields, which reduces the Einstein-Euler equations to a $1+1$-dimensional problem with periodic boundary conditions.

The article is organized as follows: the derivation of a first order formulation of the Gowdy symmetric Einstein-Euler equations that is suitable for numerical implementation and constructing solutions globally to the future is carried out in Section \ref{sec:EinsteinEulerderivation}. In Section \ref{sec:FLRWsoln}, we derive the FLRW background solutions that we perturb. Finally, in Section \ref{sec:NumericalResults}, we discuss our numerical setup and results.

\section{Einstein-Euler Equations}
\label{sec:EinsteinEulerderivation}

\subsection{Einstein-Euler equations with Gowdy symmetry}
\label{sec:EinsteinEulerderivationGowdy}

The Einstein-Euler equations\footnote{Our indexing conventions are as follows: lower case Latin letters, e.g. $i,j,k$,
will label spacetime coordinate indices that run from $0$ to $3$ while upper case Latin letters, e.g. $I,J,K$, will label spatial coordinate indices that run from
$1$ to $3$.} for a perfect fluid with a positive cosmological constant are given by 
\begin{align}
\label{eqn:Einstein1}
    G_{ij}+\Lambda g_{ij} &= T_{ij}, \;\; \Lambda > 0, \\
    \label{eqn:Tij_divergence}
    \nabla^{i}T_{ij} &= 0,
\end{align}
where
\begin{align}
\label{eqn:perfectfluidtensor}
    T_{ij} = (\rho+p)v_{i}v_{j} + pg_{ij}
\end{align}
is the stress-energy tensor of a perfect fluid, $v_{i}$ is the fluid four-velocity normalised by $g^{ij}v_{i}v_{j}=-1$, and we assume that the fluid's proper energy density, $\rho$, and pressure, $p$, are related via the linear equation of state 
\begin{align*}
    p = K\rho.
\end{align*}
Here, the constant $K$ is the square of the sound speed and in order to ensure that the speed of sound is less than or equal to the speed of light, we will always assume that $0\leq K\leq 1$. 

As discussed in the introduction, we restrict our attention to solutions of the Einstein-Euler equations with a Gowdy symmetry \cite{chrusciel1990,gowdy1974}. 
We do this by considering Gowdy metrics in areal coordinates on $\Rbb_{>0} \times \mathbb{T}^{3}$ that are of the form
\begin{align}
\label{eqn:gowdy1}
    g = e^{2\bar{\eta}-\bar{U}}(-\bar{\alpha}dt\otimes dt + d\theta \otimes d\theta) + e^{2\bar{U}}(dy+A dz)\otimes (dy+Adz)+e^{-2\bar{U}}t^{2}dz\otimes dz,
\end{align}
where the functions $\bar{\eta}$, $\bar{U}$, $\bar{\alpha}$, and $A$ depend only on $(t,\theta)\in \Rbb_{>0}\times \Tbb$. Here, we take $\theta$ to be a periodic coordinate on the 1-torus $\Tbb$ obtained by identifying the ends of the interval $[0,2\pi]$. In practice, this means that $\theta$ is a Cartesian coordinate on $\Rbb$ and that the functions  $\bar{\eta}$, $\bar{U}$, $\bar{\alpha}$, and $A$ are all $2\pi$-periodic in $\theta$. Moreover, as we are only interested in solutions in the expanding direction, i.e.\ towards the future, we will only concern ourselves with time intervals of the form $t\in [t_0,\infty)$ for some $t_0>0$.

In order to facilitate the numerical construction of solutions near timelike infinity, we first transform the metric variables via
\begin{align*}
\bar{\alpha}= \frac{e^{2\alpha}}{4t^3}, \quad
\bar{U} = U+\frac{1}{2}\log(t) \AND 
\bar{\eta} = \eta +\log(t),
\end{align*}
which allows us to express the Gowdy metric \eqref{eqn:gowdy1} as
\begin{align}\label{eqn:gowdy2}
g = t\Bigl(e^{2(\eta-U)}\Bigl(-\frac{e^{2\alpha}}{4t^3}dt\otimes dt+d\theta \otimes d\theta\Bigr)+e^{2U}(dy+Adz)\otimes (dy+Adz) +e^{-2U}dz\otimes dz\Bigr).
\end{align}
To procced, we compactify the time interval from $[1,\infty)$ to $(0,1]$ using the change of time coordinate
\begin{align*}
t = \frac{1}{\tau^{2}},
\end{align*}
which, after substituting into \eqref{eqn:gowdy2},
yields
\begin{align}
\label{eqn:gowdycompact}
g = \frac{1}{\tau^{2}}\bigl(e^{2(\eta-U)}(-e^{2\alpha} d\tau\otimes d\tau+d\theta \otimes d\theta)+e^{2U}(dy+Adz)\otimes (dy+Adz)+e^{-2U}dz\otimes dz\bigr)
\end{align}
where now $\tau \in (0,1]$ and the functions $\eta, U, \alpha$, and $A$ depend on $(\tau,\theta)$ and are $2\pi$-periodic in $\theta$. It should be noted that, due to our conventions, future timelike infinity is located at $\tau = 0$ in the direction of \textit{decreasing} $\tau$. As a result, we require $v^{0}<0$ to ensure that the four-velocity $v^\mu$ is future oriented with respect to the original time orientation.

Next, we turn to expressing the Einstein-Euler system \eqref{eqn:Einstein1}-\eqref{eqn:Tij_divergence} in a Gowdy-symmetric form suitable for numerical implementation.
To do so, we express the Einstein equations as a first order system and choose appropriate variables to formulate the Euler equations. The details of the derivation are presented in the following two sections. 

\subsection{A first order formulation of the Einstein equations}

\label{sec:EinsteinEquations}
In Gowdy symmetry, the fluid four-velocity only has two non-zero components\footnote{This follows from choosing coordinates where the two Killing vectors are given by $\del_{y}$ and $\del_{z}$, see \cite{LeFlochRendall:2011}.} and can be expressed as
\begin{equation} \label{v-Gowdy}
v=v_0 d\tau  + v_1 d\theta.
\end{equation}
Using this, we find after a short calculation that the non-zero components of the stress-energy tensor are given by
\begin{equation}
  \begin{aligned}
&T_{00} = (K+1)\rho(v_{0})^{2}-\frac{K\rho e^{2\eta-2U+2\alpha}}{\tau^{2}}, \;\; T_{01} = (K+1)\rho v_{0}v_{1}, \\
&T_{11} = (K+1)\rho(v_{1})^{2}+\frac{K\rho e^{2(\eta-U)}}{\tau^{2}}, \;\; T_{22} = \frac{K\rho e^{2U}}{\tau^{2}},\\ 
&T_{23} = \frac{K\rho Ae^{2U}}{\tau^{2}}, \;\; T_{33}=\frac{K\rho(e^{2U}A^{2}+e^{-2U})}{\tau^{2}}.
  \end{aligned} \label{eqn:T_components}
\end{equation}
With the help of these expressions and the Gowdy metric \eqref{eqn:gowdycompact}, 
a straightforward calculation shows that the Einstein equations \eqref{eqn:Einstein1} in Gowdy symmetry take the form of three wave equations 
\begin{align}
\del_{\tau\tau}A =& \frac{1}{\tau}e^{-4U}(-2e^{2\eta+2\alpha}A\tau T_{22}+2e^{2\eta+2\alpha}\tau T_{23}+4e^{4U+2\alpha}\tau\del_{\theta}A\del_{\theta}U+e^{4U+2\alpha}\tau\del_{\theta}A\del_{\theta}\alpha \nonumber \\
\label{eqn:Awave}
&+e^{4U+2\alpha}\tau\del_{\theta\theta}A+2e^{4U}\del_{\tau}A-4e^{4U}\tau\del_{\tau}A\del_{\tau}U+e^{4U}\tau\del_{\tau}A\del_{\tau}\alpha), \\ \nonumber \\
\del_{\tau\tau}U =& \frac{-1}{2\tau}e^{-4U}(-e^{2\eta+2\alpha}\tau T_{22}+e^{4U+2\eta+2\alpha}A^{2}\tau T_{22}-2e^{4U+2\eta+2\alpha}A\tau T_{23}+e^{4U+2\eta+2\alpha}\tau T_{33}+e^{8U+2\alpha}\tau(\del_{\theta}A)^{2} \nonumber \\ 
\label{eqn:Uwave}
&-2e^{4U+2\alpha}\tau\del_{\theta}U\del_{\theta}\alpha-2e^{4U+2\alpha}\tau\del_{\theta\theta}U-e^{8U}\tau(\del_{\tau}A)^{2}-4e^{4U}\del_{\tau}U-2e^{4U}\tau \del_{\tau}U\del_{\tau}\alpha), \\ \nonumber \\
\del_{\tau\tau}\eta =& \frac{e^{-2U}}{4\tau^{2}}(-12e^{2U}+4e^{2\eta+2\alpha}\Lambda-4e^{2U+2\eta+2\alpha}A^{2}\tau^{2}T_{22}+8e^{2U+2\eta+2\alpha}A\tau^{2}T_{23}-4e^{2U+2\eta+2\alpha}\tau^{2}T_{33} \nonumber \\
&-e^{6U+2\alpha}\tau^{2}(\del_{\theta}A)^{2}+4e^{2U+2\alpha}\tau^{2}(\del_{\theta}U)^{2}+4e^{2U+2\alpha}\tau^{2}\del_{\theta}\eta\del_{\theta}\alpha+4e^{2U+2\alpha}\tau^{2}(\del_{\theta}\alpha)^{2}+4e^{2U+2\alpha}\tau^{2}\del_{\theta\theta}\eta\nonumber \\
\label{eqn:Etawave}
&+4e^{2U+2\alpha}\tau^{2}\del_{\theta\theta}\alpha+e^{6U}\tau^{2}
(\del_{\tau}A)^{2}+8e^{2U}\tau\del_{\tau}U-4e^{2U}\tau^{2}(\del_{\tau}U)^{2}-8e^{2U}\tau\del_{\tau}\alpha+4e^{2U}\tau^{2}\del_{\tau}\eta\del_{\tau}\alpha),
\end{align}
and three first order equations 
\begin{align}
\label{eqn:Alphaevo}
\del_{\tau}\alpha =& \frac{-1}{2\tau}e^{-2U}(6e^{2U}-2e^{2\eta+2\alpha}\Lambda-e^{2U}\tau^{2}T_{00}+e^{2U+2\alpha}\tau^{2}T_{11}),  \\
\label{eqn:Etaevo}
\del_{\tau}\eta =& \frac{-e^{-2U}}{8\tau}(-12e^{2U}+4e^{2\eta+2\alpha}\Lambda+4e^{2U}\tau^{2}T_{00}+e^{6U+2\alpha}\tau^{2}(\del_{\theta}A)^{2}+4e^{2U+2\alpha}\tau^{2}(\del_{\theta}U)^{2} \nonumber \\
&+e^{6U}\tau^{2}(\del_{\tau}A)^{2}-8e^{2U}\tau\del_{\tau}U+4e^{2U}\tau^{2}(\del_{\tau}U)^{2}), \\
\label{eqn:Eta_derivconstraint}
 \del_{\theta}\eta =& \frac{1}{4}(-2\tau T_{01}+4\del_{\theta}U-4\del_{\theta}\alpha-e^{4U}\tau\del_{\theta}A\del_{\tau}A-4\tau\del_{\theta}U\del_{\tau}U).
\end{align}
In particular, \eqref{eqn:Etaevo} and \eqref{eqn:Eta_derivconstraint} are the Hamiltonian and momentum constraints, respectively.

In practice, either \eqref{eqn:Etawave} or \eqref{eqn:Etaevo} can be used as an evolution equations for $\eta$, however only one is needed for our numerical scheme. In this article, we use \eqref{eqn:Etaevo}. This has the benefit of enforcing the Hamiltonian constraint at every time step and it does not require solving a second order equations for $\eta$. Moreover, because we use \eqref{eqn:Etaevo} to evolve $\eta$, we can view \eqref{eqn:Etawave} as a constraint equation that can be used to verify our numerical results.

Next, introducing the first order variables
\begin{align}
\label{eqn:firstordervariablesA_U}
A_{0} &= \del_{\tau}A, \quad A_{1} = e^{\alpha}\del_{\theta}A, \quad U_{0} = \del_{\tau}U, \quad U_{1} = e^{\alpha}\del_{\theta}U,
\end{align}
we can, with the help of \eqref{eqn:T_components}, express the wave equations \eqref{eqn:Awave}-\eqref{eqn:Uwave} for $A$ and $U$ in first order form as
\begin{align}
\label{eqn:Asymhyp}
\del_{\tau}\begin{pmatrix} A_{0} \\ A_{1} \end{pmatrix} + \begin{pmatrix} 0 & -e^{\alpha} \\ -e^{\alpha} & 0 \end{pmatrix} \del_{\theta}\begin{pmatrix} A_{0} \\ A_{1} \end{pmatrix}  -\alpha_{0}\begin{pmatrix} A_{0} \\ A_{1} \end{pmatrix} =& \begin{pmatrix} \frac{1}{\tau}(4\tau A_{1}U_{1}+2A_{0}-4\tau A_{0}U_{0}) \\ 0 \end{pmatrix}, \\
\label{eqn:Usymhyp}
\del_{\tau}\begin{pmatrix} U_{0} \\ U_{1} \end{pmatrix} + \begin{pmatrix} 0 & -e^{\alpha} \\ -e^{\alpha} & 0 \end{pmatrix} \del_{\theta}\begin{pmatrix} U_{0} \\ U_{1} \end{pmatrix}  -\alpha_{0}\begin{pmatrix} U_{0} \\ U_{1} \end{pmatrix} =& \begin{pmatrix} \frac{-1}{2\tau}(e^{4U}\tau A_{1}^{2}-e^{4U}\tau(A_{0})^{2}-4U_{0})\\ 0 \end{pmatrix},
\end{align}
while the remaining Einstein equations are given by
\begin{align}
\label{eqn:alphaevo2}
\alpha_{0} :=&\; \del_{\tau}\alpha = \frac{-e^{-2U}}{2\tau}(6e^{2U}-2e^{2\eta+2\alpha}\Lambda+e^{2U}\tau^{2}(K+1)\rho (v_{1}^{2}e^{2\alpha}-v_{0}^{2})+2K\rho e^{2(\eta+\alpha)}), \\
\label{eqn:etaevo2}
 \del_{\tau}\eta =& \frac{-1}{8\tau}\Bigg(-12+4e^{2(\eta+\alpha-U)}(\Lambda-K\rho)+\tau\bigg(e^{4U}\tau(A_{1}^{2}+A_{0}^{2})-8U_{0}+4\tau\Big((1+K)\rho v_{0}^{2}+U_{1}^{2}+U_{0}^{2}\Big)\bigg)\Bigg), \\
\label{eqn:etawave2}
\del_{\tau\tau}\eta =&\; \frac{1}{4} e^{4 U} (A_{0}^{2}-A_{1}^{2})+\frac{(\Lambda -K \rho ) e^{2 (\alpha
   -U+\eta)}+\tau U_{0} (2-\tau U_{0})+\tau \alpha_{0} (\tau
   \del_{\tau}\eta-2)-3}{\tau^2} \nonumber \\
   &+U_{1}^{2}+e^{2 \alpha } (\del_{\theta\theta}\alpha +\del_{\theta}\alpha 
   (\del_{\theta}\alpha +\del_{\theta}\eta)+\del_{\theta\theta}\eta), \\
\label{eqn:Aevo}
\del_{\tau}A =&\; A_{0}, \\
\label{eqn:Uevo}
\del_{\tau}U =&\; U_{0},\\
\label{eqn:Eta_derivconstraint2}
 \del_{\theta}\eta =& \frac{1}{4}\Big(-2(1+K)\rho v_{0}v_{1}\tau-4\del_{\theta}\alpha-e^{4U}\tau A_{0}\del_{\theta}A +\del_{\theta}U(4-4\tau U_{0})\Big),
\end{align}
where \eqref{eqn:etawave2} and \eqref{eqn:Eta_derivconstraint2} are constraint equations.

\subsection{A first order formulation of the Euler equations}
\label{sec:EulerEquations}
Contracting the Euler equations \eqref{eqn:Tij_divergence} with the fluid four-velocity $v^{j}$ and the projection operator $L^{j}_{J} = \delta^{j}_{J} - \frac{v_{J}}{v_{0}}\delta^{j}_{0}$, respectively, we find with the help of the normalization condition $v_i v^i =-1$ that the Euler equations can be expressed as\footnote{Note this formulation of the Euler equations was first derived in \cite{Oliynyk:CMP_2015}.}
\begin{align}
A^{i}\nabla_{i}\begin{pmatrix}\rho \\ v_{K} \end{pmatrix} = \begin{pmatrix}0\\0\end{pmatrix}
\label{Euler-A}
\end{align}
where the coefficient matrix $A^{i}$ is given by
\begin{align*}
A^{i} = \begin{pmatrix} \frac{K}{\rho+p}v^{i} & Kg^{Kl}(\delta^{i}_{l}-\frac{v_{l}}{v^{0}}g^{i0}) \\ 
Kg^{Jl}(\delta^{i}_{l}-\frac{v_{l}}{v^{0}}g^{i0}) & (\rho +p)g^{Jl}g^{Ka}(g_{al}-\frac{2}{v^{0}}v_{(a}\delta^{0}_{l)}+\frac{v_{l}v_{a}}{(v^{0})^{2}}g^{00})v^{i} \end{pmatrix}
\end{align*}
and $\nabla_i v_K = \del_{i}v_K - \Gamma_{iK}^j v_j$. 

We now note by \eqref{v-Gowdy} that in Gowdy symmetry $\nabla_i v_{K}$ can be expressed as
\begin{align} \label{CDv-iK}
\nabla_{i}v_{K} = \del_{i}v_{K} - \frac{1}{2}\Big(g^{00}(\del_{K}g_{i0}-\del_{0}g_{iK})v_{0}+g^{11}(\del_{i}g_{K1}+\del_{K}g_{i1}-\del_{1}g_{iK})v_{1}\Big),
\end{align}
and that the normalisation condition $v_i v^i=-1$ is given by
\begin{align*}
g^{00}(v_{0})^{2} + g^{11}(v_{1})^{2} =  -1,
\end{align*}
which can be solved for $v_{0}$ to obtain 
\begin{align}
\label{eqn:v0identity}
v_{0} = \sqrt{-g_{00}-g_{00}g^{11}(v_{1})^{2}}.
\end{align}
Then, with the help of \eqref{v-Gowdy}, \eqref{CDv-iK} and \eqref{eqn:v0identity},
we find following a straightforward calculation
that the Euler equations \eqref{Euler-A} can 
be written as
\begin{align} \label{Euler-B}
B^{0}\del_{0}V + B^{1}\del_{1}V = F, 
\end{align}
where
\begin{align*}
V =& \begin{pmatrix}\rho \\ v_{1} \end{pmatrix},\\
B^{0} =& \begin{pmatrix} \frac{K}{\rho+K\rho}\left(g_{11}+(v_{1})^{2}\right) & Kv_{1} \\ Kv_{1} & \rho+K\rho\end{pmatrix}, \\
B^{1} =& \begin{pmatrix} \frac{K}{\rho+K\rho}v_{1} & K \\ K & (\rho+K\rho)\frac{v_{1}}{g_{11}+(v_{1})^{2}}\end{pmatrix},
\end{align*}
and
\begin{align*}
&F = \frac{1}{2}(-v_{0})\begin{pmatrix} K\left(2g^{11}\del_{1}g_{11}-g^{ab}\del_{1}g_{ab}\right)v_{1} \\(\rho+K\rho)\left(\frac{(v_{1})^{2}}{g_{11}+(v_{1})^{2}}g^{11}\del_{1}g_{11} - g^{00}\del_{1}g_{00}\right)\end{pmatrix} +\frac{K}{2} \begin{pmatrix} (v_{1})^{2}g^{11}\del_{0}g_{11}-\big(g_{11}+(v_{1})^{2}\big)g^{IK}\del_{0}g_{IK} \\ 0 \end{pmatrix}.
\end{align*}

To proceed, we define re-scaled Gowdy fluid variables $(\rhot,\tilde{v}_{1})$ via
\begin{align}
\label{eqn:vt1change}
v_{1} &= \tau^{-\mu-1}\tilde{v}_{1} \\
\label{eqn:rhotdef}
\rho &= \tau^{3(1+K)}\rhot,
\end{align}
where $\mu = \frac{3K-1}{1-K}$. The particular powers of $\tau$ in the above definitions are  chosen to remove the expected leading order behavior in $\tau$. Now, in order to express the Euler equations \eqref{Euler-B} in terms of these new variables, we differentiate \eqref{eqn:vt1change}-\eqref{eqn:rhotdef} to obtain the identities
\begin{align*}
\del_{0}\begin{pmatrix}\rho \\ v_{1} \end{pmatrix}&= P\del_{0}\begin{pmatrix}\rhot \\ \tilde{v}_{1}\end{pmatrix} +Z,\\ 
\del_{1}\begin{pmatrix}\rho \\ v_{1} \end{pmatrix}&= P\del_{1}\begin{pmatrix}\rhot \\ \tilde{v}_{1}\end{pmatrix},
\end{align*}
where 
\begin{equation*}
P = \begin{pmatrix}\tau^{3(1+K)} & 0 \\ 0 & \tau^{-\mu-1} \end{pmatrix} \AND
Z = \begin{pmatrix} 3(1+K)\rhot\tau^{2+3K} \\  (-\mu-1) \tilde{v}_{1}\tau^{-\mu-2} \end{pmatrix}.
\end{equation*}
Using these identities, it is straightforward to verify that the Euler equations \eqref{Euler-B} can be expressed as
\begin{align}
\label{eqn:Eulertilde}
\tilde{B}^{0}\del_{0}\tilde{V} + \tilde{B}^{1}\del_{1}\tilde{V} = \tilde{F},
\end{align}
where the matrices $\tilde{B}^{i}, \tilde{V},$ and $\tilde{F}$ are defined\footnote{Here T denotes the transpose of a matrix.} by
\begin{align*}
\tilde{B}^{0} = P^{\text{T}}B^{0}P, \quad 
\tilde{B}^{1} = P^{\text{T}}B^{1}P, \quad
\tilde{V} = \begin{pmatrix}\rhot \\ \tilde{v}_{1}\end{pmatrix} \AND
\tilde{F} = P^{\text{T}}(F - B^{0}Z),
\end{align*}
respectively.

\subsection{The complete evolution system}
Combining \eqref{eqn:Eulertilde} with \eqref{eqn:Asymhyp}, \eqref{eqn:Usymhyp}, \eqref{eqn:alphaevo2}, \eqref{eqn:etaevo2}, \eqref{eqn:Aevo}, and \eqref{eqn:Uevo} 
yields a closed set of evolution equations
that we will solve numerically. These
equations can be expressed in matrix form as
\begin{align}
\label{eqn:EinsteinEuler1}
\begin{pmatrix} \mathbb{I} & 0 & 0 \\ 0 & \mathbb{I} & 0 \\ 0 & 0 & \tilde{B}^{0} \end{pmatrix} \del_{\tau} \begin{pmatrix} \textbf{A} \\ \textbf{U} \\ \tilde{\textbf{V}} \end{pmatrix} +\begin{pmatrix} \bar{B}^{1}  & 0 & 0 \\ 0 & \bar{B}^{1}  & 0 \\ 0 & 0 & \tilde{B}^{1} \end{pmatrix} \del_{\theta} \begin{pmatrix} \textbf{A} \\ \textbf{U} \\ \tilde{\textbf{V}} \end{pmatrix} &= \begin{pmatrix} \alpha_{0} \mathbb{I} & 0 & 0 \\ 0 & \alpha_{0} \mathbb{I} & 0 \\ 0 & 0 & 0 \end{pmatrix} \begin{pmatrix} \textbf{A} \\ \textbf{U} \\ \tilde{\textbf{V}} \end{pmatrix} + \begin{pmatrix} F_{A} \\ F_{U} \\ F_{\tilde{V} }\end{pmatrix}, \\
\label{eqn:EinsteinEuler2}
\del_{\tau} \begin{pmatrix} \alpha \\ \eta \\ A \\ U \end{pmatrix} &= \begin{pmatrix} F_{\alpha} \\ F_{\eta} \\ A_{0} \\ U_{0} \end{pmatrix}, 
\end{align}
where
\begin{align*}
\textbf{A} =& \begin{pmatrix} A_{0} \\ A_{1} \end{pmatrix}, \\
\textbf{U} =& \begin{pmatrix} U_{0} \\ U_{1} \end{pmatrix}, \\ 
\tilde{\textbf{V}} =& \begin{pmatrix} \rhot \\ \tilde{v}_{1} \end{pmatrix},  \\
\bar{B}^{1} =& \begin{pmatrix} 0 & -e^{\alpha} \\ -e^{\alpha} & 0 \end{pmatrix},  \\ 
\tilde{B}^{0} =& \begin{pmatrix} \frac{K \tau^{3 K+1} \left(e^{2 \eta-2 U}+\tilde{v}_{1}^2 \tau^{-2 \mu }\right)}{(K+1) \rhot} & K \tilde{v}_{1} \tau^{3 K-2 \mu +1} \\ K \tilde{v}_{1} \tau^{3 K-2 \mu +1} & (K+1) \rhot \tau^{3 K-2 \mu +1} \\\end{pmatrix},  \\
\tilde{B}^{1} =& \begin{pmatrix}-\frac{K \tilde{v}_{1} \tau^{3 K-\mu +1}e^{\alpha } \sqrt{ \tilde{v}_{1}^2 \tau^{-2 \mu }+e^{2 \eta-2 U}}}{(K+1) \rhot} & -K \tau^{3K-\mu +1}e^{ \alpha } \sqrt{ \tilde{v}_{1}^2 \tau^{-2 \mu }+e^{2 \eta-2 U}} \\-K \tau^{3 K-\mu +1} e^{\alpha }\sqrt{ \tilde{v}_{1}^2 \tau^{-2 \mu }+e^{2 \eta-2 U}} & -\frac{e^{\alpha}(1+K)\rhot \tilde{v}_{1}\tau^{3K-3\mu+1}}{\sqrt{e^{-2U+2\eta}+\tilde{v}_{1}^{2}\tau^{-2\mu}}}\\\end{pmatrix},  \\
F_{A} =& \begin{pmatrix} \frac{1}{\tau}\left(4\tau A_{1}U_{1}+2A_{0}-4\tau A_{0}U_{0}\right) \\ 0 \end{pmatrix},  \\
F_{U} =& \begin{pmatrix} \frac{-1}{2\tau}\left(e^{4U}\tau A_{1}^{2}-e^{4U}\tau(A_{0})^{2}-4U_{0}\right)\\ 0 \end{pmatrix},  \\
F_{\tilde{V}} =& \begin{pmatrix}K \tau^{3 K+1} e^{2 \eta-2 U} \left(U_{0}-\del_{\tau}\eta\right)+K \tilde{v}_{1} \tau^{3 K-\mu +1} e^{\alpha }\sqrt{e^{2\eta-2 U}+\tilde{v}_{1}^2 \tau^{-2 \mu }}\del_{\theta}\alpha +K \mu \tilde{v}_{1}^2 \tau^{3 K-2 \mu } \\(K+1) \rhot \tau^{3 K-2 \mu } \left(\frac{\tau^{\mu +1}e^{\alpha } \sqrt{ e^{2 \eta-2 U}+\tilde{v}_{1}^2 \tau^{-2 \mu}} \left(\tau^{2 \mu } e^{2 \eta} \left(\del_{\theta}\alpha-\del_{\theta}U+\del_{\theta}\eta\right)+\tilde{v}_{1}^2 e^{2 U}\del_{\theta}\alpha \right)}{\tilde{v}_{1}^2 e^{2 U}+\tau^{2 \mu } e^{2 \eta}}+\tilde{v}_{1}(-3 K+\mu +1)\right) \\\end{pmatrix},  \\
F_{\alpha} =& \frac{\bigg(2 \Lambda -(K-1) \rhot \tau^{3 K+3}\bigg) e^{ 2 (\alpha -U+\eta)}-6}{2 \tau},  \\
F_{\eta} =& \frac{1}{8}e^{-2U}\tau^{-1-2\mu}\Bigg(-4e^{2(U+\alpha)}(1+K)\rhot \tilde{v}_{1}^{2}\tau^{3+3K}-\tau^{2\mu}\bigg(4\left(-3e^{2U}+e^{2(\eta+\alpha)}(\Lambda+\rhot \tau^{3+3K})\right) \nonumber \\
&+e^{2U}\tau\Big(e^{4U}\tau(A_{1}^{2}+A_{0}^{2})-8\del_{\tau}U+4\tau\big(U_{1}^{2}+U_{0}^{2}\big)\Big)\bigg)\Bigg).
\end{align*}

\begin{rem}
While the form of the equations \eqref{eqn:EinsteinEuler1}-\eqref{eqn:EinsteinEuler2} is suitable for numerical implementation, it is not immediately obvious that the system has a well-posed initial value problem. By re-writing the equations as a symmetric hyperbolic system we can ensure this is the case. The Euler equations prove to be the only impediment to this goal, in particular the derivatives of metric functions in the source term $F_{\tilde{V}}$ necessitate the use of new variables. By slightly modifying the process in \cite{GrubicLeFloch:2013} and introducing the scalar velocity $v = \frac{v^{1}}{e^{\alpha}v^{0}}$, a new metric variable $\nu = \eta +\alpha$, and a modified density variable $\gamma = e^{\nu-U}\rho$, it is possible to write the Euler equations in the form 
\begin{align}
\label{eqn:Remark_1}
C^{0}\del_{\tau}\begin{pmatrix}\gamma \\ v \end{pmatrix}+C^{1}\del_{\theta}\begin{pmatrix}\gamma \\ v \end{pmatrix} = G
\end{align}
where $G$ only contains derivatives of $U$ and $A$ which can be expressed in terms of the first order variables defined earlier \eqref{eqn:firstordervariablesA_U}. Multiplying \eqref{eqn:Remark_1} on the left by $P(C^{0})^{-1}$ for an appropriate symmetric matrix $P$, it is then possible to put the Euler equations in symmetric hyperbolic form. Finally, by replacing all remaining terms in \eqref{eqn:EinsteinEuler1}-\eqref{eqn:EinsteinEuler2} with the new variables, the Einstein-Euler equations in Gowdy symmetry can be cast in symmetric hyperbolic form.
\end{rem}

\section{FLRW Solutions}
\label{sec:FLRWsoln}
Before we can choose appropriate initial data for our numerical scheme, we must first identify the FLRW solutions (i.e.\ spatially homogeneous and isotropic) that we wish to perturb. Recalling the form of the Gowdy metric \eqref{eqn:gowdycompact},
we observe that a FLRW metric is 
obtained by setting $U=A=\eta = 0$
and assuming that the remaining metric function $\alpha$ only depends on $\tau$. For the Gowdy fluid variables $\rhot$ and $\tilde{v}_{1}$, spatial homogeneity and isotropy requires that $\tilde{v}_{1}=0$ and that
$\rhot$ also only depends on $\tau$. From these considerations, we conclude via \eqref{eqn:EinsteinEuler1} and \eqref{eqn:EinsteinEuler2} that FLRW solutions of the Einstein-Euler equations are obtained from solving 
\begin{align}
\label{eqn:rhohomog}
\del_{\tau}\rhot &= 0, \\
\label{eqn:alphahomog}
\del_{\tau}\alpha &= \frac{\left(2 \Lambda -(K-1) \rhot \tau^{3 K+3}\right) e^{2\alpha}-6}{2 \tau},  \\
\label{eqn:etahomog}
\del_{\tau}\eta&=0=\frac{(\Lambda+\rhot \tau^{3K+3})e^{2\alpha}-3}{2\tau}.
\end{align}

Now, by \eqref{eqn:etahomog}, we observe that
\begin{align*}
\alpha = \frac{1}{2}\ln\left(\frac{3}{\Lambda+\rhot \tau^{3K+3}}\right),
\end{align*}
which we note automatically satisfies \eqref{eqn:alphahomog}. Furthermore,
we find from \eqref{eqn:rhotdef} and \eqref{eqn:rhohomog} that 
\begin{align*}
\rho = \rho_{c}\tau^{3(1+K)},
\end{align*}
where $\rho_{c} \in \mathbb{R^{+}}$ is a freely specifiable constant. From this, we deduce that the FLRW solutions of the Einstein-Euler equations are given by
\begin{equation} \label{eqn:solnhomog}
\begin{aligned}
g &= \frac{1}{\tau^{2}}\Bigl(-\frac{3}{\Lambda+\rho_{c}\tau^{3(1+K)}}\dd \tau^{2} +\dd\theta^{2}+\dd y^{2} + \dd z^{2}\Bigr), \\ 
\rho &= \rho_{c}\tau^{3(1+K)}, \\
v &= \Bigl(\frac{3}{\Lambda+\rho_{c}\tau^{3(1+K)}}\Bigr)^{\frac{1}{2}}d\tau.
\end{aligned} 
\end{equation}

\begin{rem} 
Expressing the momentum constraint equations \eqref{eqn:Eta_derivconstraint} in terms of $\tilde{v}_{1}$ and $\rhot$,
we observe that
\begin{align} 
\label{eqn:momentumconstrainthomogexp}
&\del_{\theta}\eta = -\frac{1}{2}(1+K)\rhot \tilde{v}_{1}\tau^{2+3K-\mu}e^{\alpha}\sqrt{(e^{2\eta-2U}+\tilde{v}_{1}^{2}\tau^{-2\mu})} - \del_{\theta}\alpha - \frac{1}{4}e^{4U}\tau A_{0}\del_{\theta}A + \del_{\theta}U(1-\tau U_{0}).
\end{align}
Since all spatial derivatives would vanish on a spatially homogeneous, but not necessarily isotropic, solution, it follows from the positivity of the density, i.e.\ $\rhot>0$ everywhere, that $\tilde{v}_{1}=0$ must be satisfied for all spatially homogeneous solutions. This, in particular, shows that self-gravitating versions of the non-isotropic spatially homogeneous fluid solutions of the type considered in \cite{Marshalloliynyk:2022}, known as \emph{tilted} solutions, are incompatible with Gowdy symmetry. As it turns out,  tilted solutions require a non-trivial spatial topology; see \cite{GoliathEllis:1999}. We will report on the nonlinear stability of tilted solutions in a separate article. 
\end{rem}

\section{Numerical Results}
\label{sec:NumericalResults}

\subsection{Numerical Setup}

In the numerical setup that we use to solve \eqref{eqn:EinsteinEuler1}-\eqref{eqn:EinsteinEuler2}, the computational spatial domain is $[0,2\pi]$ with periodic boundary conditions that is discretised using an equidistant grid with $N$ grid points. Spatial derivatives are discretised using $2^{\text{nd}}$ order central finite differences and time integration is performed using a standard $2^{\text{nd}}$ order Runge-Kutta method (\textit{Heun's Method}). As a consequence, our code is second order accurate. We also enforce the CFL condition to ensure convergence. In this case we have used the tightened 4/3 CFL condition for Heun's Method which is discussed in \cite{schneider:hal-01307287}. 

\subsubsection{Initial Data}
The choice of initial data is not completely trivial as we must satisfy the Hamiltonian \eqref{eqn:etaevo2} and momentum \eqref{eqn:Eta_derivconstraint2} constraints initially. The Hamiltonian constraint \eqref{eqn:etaevo2} is enforced at every time-step, as we use it as an evolution equation for $\eta$. Consequently, we only need to ensure our choice of initial data satisfies the momentum constraint \eqref{eqn:Eta_derivconstraint2}. Additionally, we must satisfy the constraints \eqref{eqn:firstordervariablesA_U} that arise from the definition of the first order variables $A_{1}$ and $U_{1}$. Our choice of initial data \eqref{eqn:numericalID} ensures all these constraints are satisfied initially.

As discussed in the introduction, the main aim of this article is to determine whether the fractional density gradient blows up when the fluid is coupled to the gravitational field in the same way as we observed in the fixed background spacetime case \cite{Marshalloliynyk:2022}. Hence, we must choose initial data so that the fluid's spatial velocity vanishes somewhere on the domain initially.

For Gowdy symmetry, this amounts to the initial data for $\tilde{v}_{1}$ vanishing somewhere on the initial hypersurface. 
Moreover, since $\tilde{v}_{1} = 0$ for the FLRW solutions, we must select initial data so that $\tilde{v}_{1}$ is everywhere close to zero on the initial hypersurface in order for it to represent a small perturbation of FLRW initial data. 
In our numerical simulations, we satisfy these constraints on the initial data for $\tilde{v}_{1}$ by using sinusoidal functions with a small amplitude parameter called $a$ below. In particular, our initial data for $\tilde{v}_{1}$ crosses zero at least twice on the initial hypersurface, and we note that this initial data is essentially the same as was used in \cite{Marshalloliynyk:2022}.

For the remainder of this article, with the exception of Section \ref{sec:codevalidation}, we employ initial data of the form
\begin{equation}
\begin{aligned}
\label{eqn:numericalID}
    \mathring{\tilde{v}}_{1} &= a\sin(\theta ), \\
    \mathring{\rhot} &= \frac{1}{\frac{1}{2}(1+K)\sqrt{e^{2\mathring{\alpha}}(e^{2\mathring{\eta}-2\mathring{U}}+\mathring{\tilde{v}}_{1}^{2})}}, \\
    \mathring{\alpha} &= a\cos(\theta ) + \frac{1}{2}\log\left(\frac{3}{\Lambda+\frac{2}{(1+K)}}\right) , \\
   \mathring{\eta} &= a\sin(\theta ), \\
    \mathring{U} &= a\sin(\theta )+c, \\
    \mathring{U}_{0} &= bd, \\
    \mathring{U}_{1} &= e^{\mathring{\alpha}}\del_{\theta}\mathring{U}, \\ 
    \mathring{A} &= de^{-4c-4a\sin(\theta )}+c, \\
    \mathring{A}_{0} &= b, \\
    \mathring{A}_{1} &= e^{\mathring{\alpha}}\del_{\theta}\mathring{A},
\end{aligned}
\end{equation}
where $a$, $b$, $c$, $d$ are constants to be specified. Initial data of this form can be considered as a perturbation of FLRW initial data
provided that the constants $a$, $b$, $c$ and $d$ are chosen sufficiently close to zero. This follows from the fact that setting $a=b=c=d=0$ in \eqref{eqn:numericalID} produces FLRW initial data. If the size of the parameters $a,b,c$, $d$ are too large the system is found to become unstable almost immediately. That is, within a small amount of timesteps the variables develop steep gradients and produce numerical errors.  Throughout this article we focus exclusively on initial data with small amplitudes. In particular, all the plots in this section have been generated with $a=b=c=d=0.01$.

\subsubsection{Code Tests}
\begin{figure}[htbp]
\centering
\subfigure[Subfigure 1 list of figures text][$\tilde{v}_{1}$]{
\includegraphics[width=0.4\textwidth]{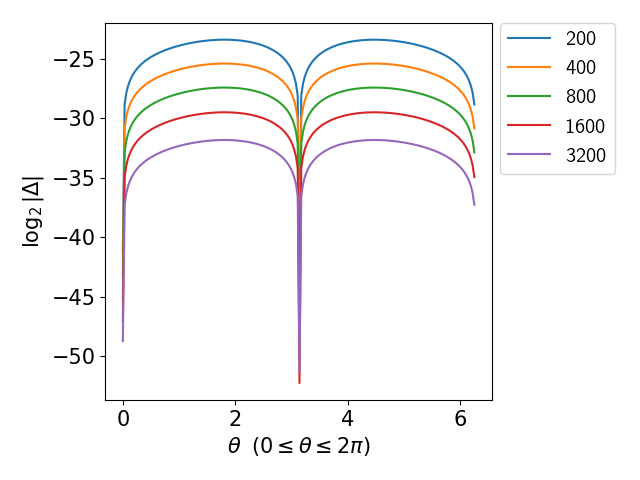}
\label{fig:subfig1}}
\subfigure[Subfigure 2 list of figures text][$\rhot$]{
\includegraphics[width=0.4\textwidth]{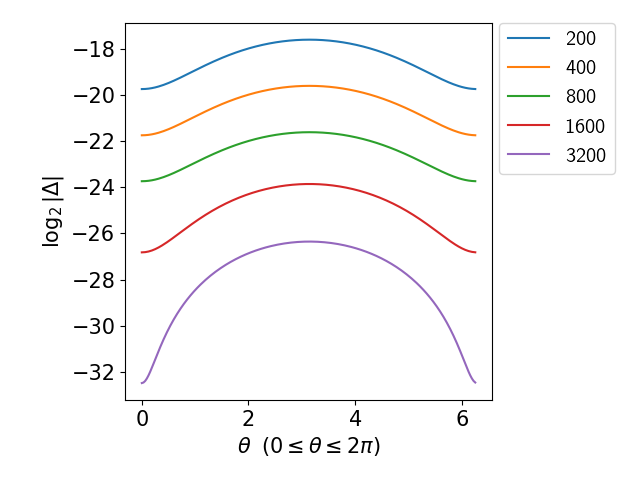}
\label{fig:subfig2}}
\caption{Convergence plots of $\tilde{v}_{1}$ and $\rhot$ at $\tau = 0.599$,  $K=0.5$, $\Lambda = 1$.}
\end{figure}

\begin{figure}[htbp]
\centering
\subfigure[Subfigure 1 list of figures text][$\log_{2}(\|C_{1}\|_{2}+\|C_{2}\|_{2}+\|C_{3}\|_{2})$ Constraint]{
\includegraphics[width=0.4\textwidth]{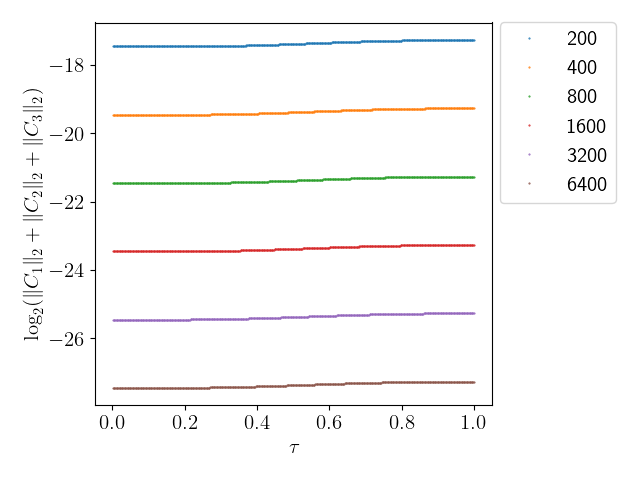}
\label{fig:subfigCsum}}
\subfigure[Subfigure 2 list of figures text][$\log_{2}(\|C_{4}\|_{2})$ Constraint]{
\includegraphics[width=0.4\textwidth]{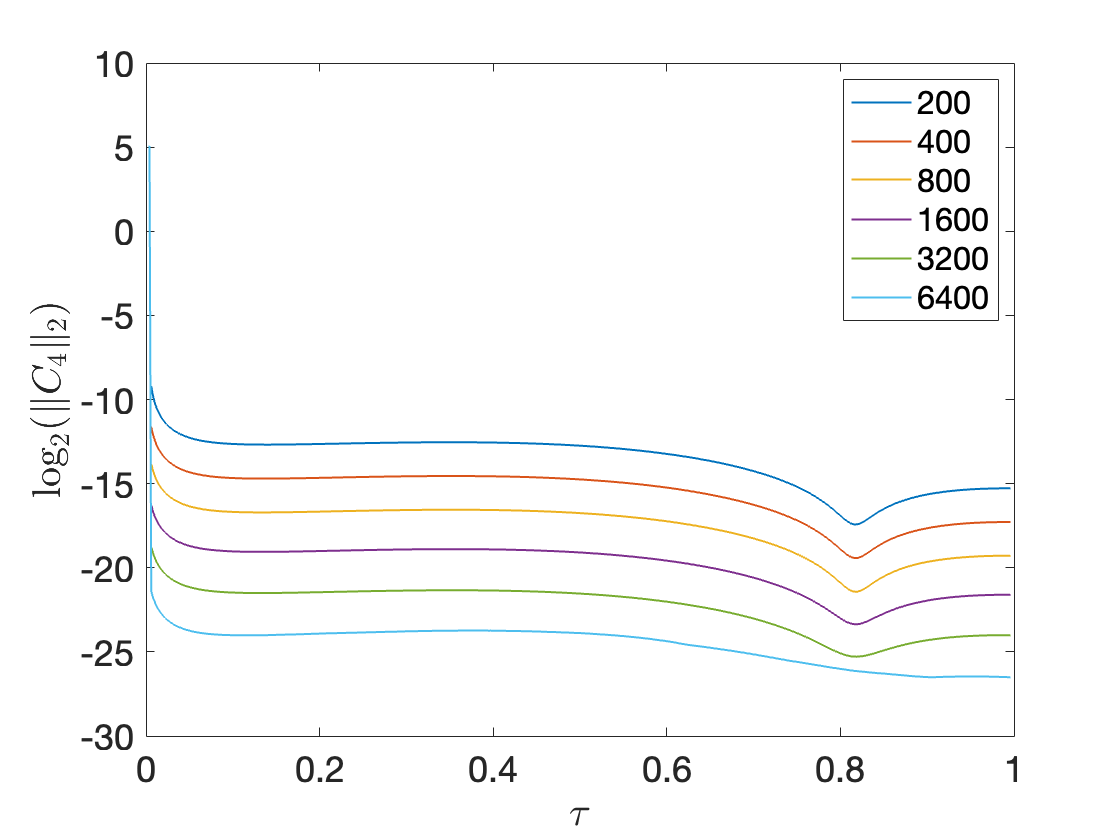}
\label{fig:subfigC4}}
\caption{Convergence plots of the constraint quantities, $K=0.5$, $\Lambda = 1$. The system was evolved until $\tau=0.002$. }
\end{figure}

We have verified the second order accuracy of our code with convergence tests involving perturbations of FLRW solutions using resolutions of $N =$ $200$, $400$, $800$, $1600$, $3200$, and $6400$ grid points. To estimate the numerical discretisation error $\Delta$ for any of our unknowns, we took the $\log_{2}$ of the absolute value of the difference between each simulation and the highest resolution run. The results for $\tilde{v}_{1}$ and $\rhot$  are shown\footnote{It should be noted that we have also performed convergence tests for all other variables and confirmed second order convergence. These plots are omitted here for brevity.} in Figures \ref{fig:subfig1}-\ref{fig:subfig2} from which the second order convergence is clear. \newline \newline
As a further check on the accuracy of the code, we can measure how much the constraints are violated during the evolution of the system. Beginning with the momentum constraint \eqref{eqn:momentumconstrainthomogexp}, we define the quantity
\begin{align}
\label{eqn:momentumconstraintquantity}
    C_{1} = -\del_{\theta}\eta -\frac{1}{2}(1+K)\rhot \tilde{v}_{1}\tau^{2+3K-\mu}e^{\alpha}\sqrt{(e^{2\eta-2U}+\tilde{v}_{1}^{2}\tau^{-2\mu})} - \del_{\theta}\alpha - \frac{1}{4}e^{4U}\tau A_{0}\del_{\theta}A + \del_{\theta}U(1-\tau U_{0}).
\end{align}
Clearly, $C_{1}=0$ means that the momentum constraint is identically satisfied. The quantity $\log_2 \norm{C_1}_{2}$ can therefore be understood as the violation error of the momentum constraint as a function of time.  In a similar manner, we can also define constraint violation quantities from the definitions of our first order variables $A_{1}$ and $U_{1}$, and from the wave equation for $\eta$, \eqref{eqn:etawave2}, as follows
\begin{align*}
    C_{2} =& A_{1}-e^{\alpha}\del_{\theta}A, \\
    C_{3} =& U_{1}-e^{\alpha}\del_{\theta}U, \\
    C_{4} =& \del_{\tau\tau}\eta -\Bigg(\frac{1}{4} e^{4 U} (A_{0}^{2}-A_{1}^{2})+\frac{(\Lambda -K \rhot \tau^{3+3K} ) e^{2 (\alpha
   -U+\eta)}+\tau U_{0} (2-\tau U_{0})+\tau \alpha_{0} (\tau
   \del_{\tau}\eta-2)-3}{\tau^2} \nonumber \\
   &+U_{1}^{2}+e^{2 \alpha } (\del_{\theta\theta}\alpha +\del_{\theta}\alpha 
   (\del_{\theta}\alpha +\del_{\theta}\eta)+\del_{\theta\theta}\eta)\Bigg).
\end{align*}
The time derivatives for $C_{4}$ are calculated numerically using a fourth order finite difference stencil for the second derivative
\begin{align}
\label{eqn:time_finitediff}
(\del_{\tau\tau}\eta)_{i,j} = \frac{-\eta_{i-2,j}+16\eta_{i-1,j}-30\eta_{i,j}+16\eta_{i+1,j}-\eta_{i+2,j}}{12(\Delta\tau)^{2}},
\end{align}
where $\eta_{i,j}$ denotes the value of $\eta$ at the i\textsuperscript{th} timestep and j\textsuperscript{th} spatial grid point and $\Delta\tau$ is the timestep size.  
We observe the expected second order convergence for the quantities $\log_{2}(\|C_{1}\|_{2}+\|C_{2}\|_{2}+\|C_{3}\|_{2})$ and $\log_{2}(\|C_{4}\|_{2})$  shown in Figures \ref{fig:subfigCsum} and \ref{fig:subfigC4}, respectively. Even though the constraints are identically satisfied at the initial time by virtue of our choice of initial data \eqref{eqn:numericalID}, we note the numerical value is not exactly zero, even at the initial time $\tau=1$, as the derivatives in $C_1$ \eqref{eqn:momentumconstraintquantity}, $C_{2}$, $C_{3}$, and $C_{4}$ are approximated by finite differences. It should also be noted that, due to our use of the stencil \eqref{eqn:time_finitediff}, the first and last two timesteps have been removed from Figure \ref{fig:subfigC4}. 

\begin{figure}
    \centering
    \includegraphics[width=0.4\textwidth]{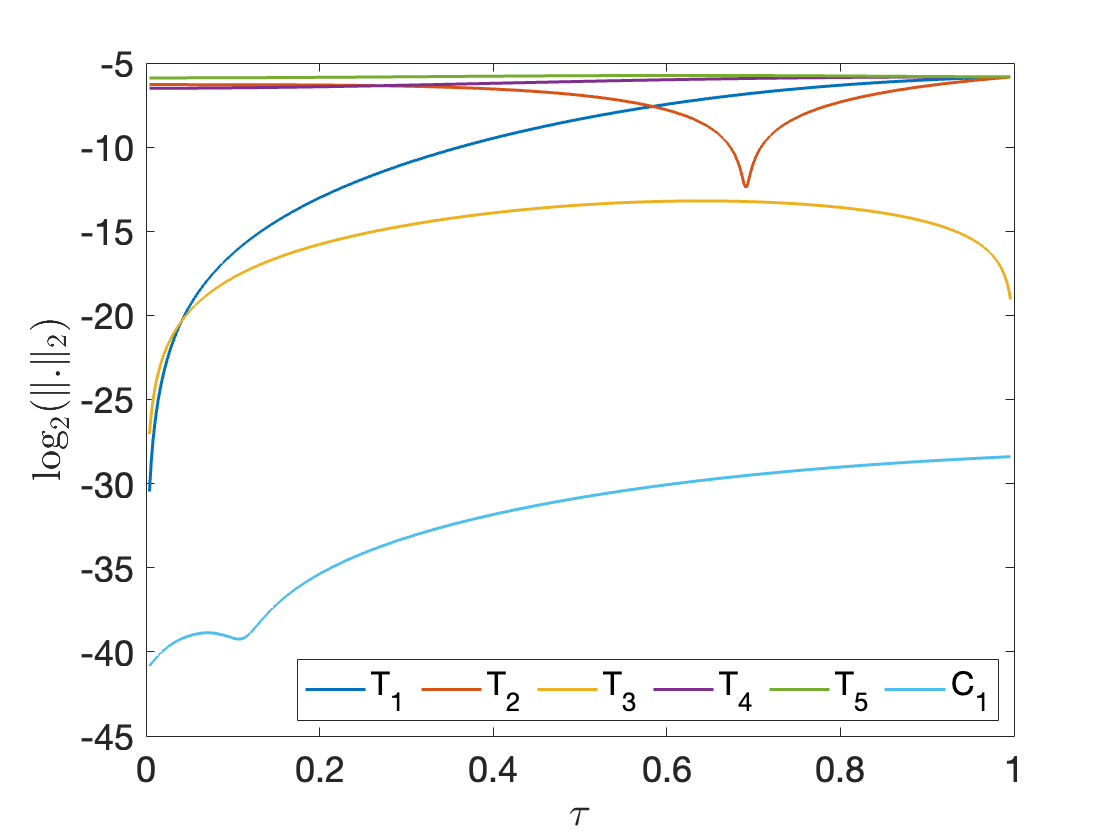}
    \caption{Comparison of the $L_{2}$ norm of the individual terms in momentum constraint and the combined constraint quantity $C_{1}$. $N=6400$, $K=0.5$, $\Lambda = 1$.}
    \label{fig:constraint_termcompare}
\end{figure}

Another measure of how well the constraints are satisfied numerically is to compare the size of each individual term in a constraint with the total constraint violation. From this we can conclude that the actual constraint violation is small (as opposed to each individual term being small). To this end we consider first $C_1$ and separate it into five terms as follows:
\begin{align}
\label{eqn:T1-constraint}
    T_{1} &= -\frac{1}{2}(1+K)\rhot \tilde{v}_{1}\tau^{2+3K\mu}e^{\alpha}\sqrt{(e^{2\eta-2U}+\tilde{v}_{1}^{2}\tau^{-2\mu})}, \\
    T_{2} &= -\del_{\theta}\alpha, \\
    T_{3} &= \frac{-e^{4U}\tau A_{0}\del_{\theta}A}{4}, \\
    T_{4} &= \del_{\theta}U(1-\tau U_{0}), \\
    \label{eqn:T5-constraint}
    T_{5} &= -\del_{\theta}\eta.
\end{align}
For the constraint violations $C_1$ to be actually small, we expect that the norm of each individual term \eqref{eqn:T1-constraint}-\eqref{eqn:T5-constraint} should be larger than the norm of the total constraint violation $C_1$ since this indicates that a cancellation among the terms in the sum is occurring. Figure~\ref{fig:constraint_termcompare} demonstrates that this cancellation is happening for $C_1$. We observe similar behaviour for the other constraints, $C_{2}$, $C_{3},$ and $C_{4}$. From these observations, we conclude that the constraints are being preserved by our numerical scheme.

\subsubsection{Code Validation}\label{sec:codevalidation}
\begin{figure}[htbp]
\centering
\subfigure[Subfigure 1 list of figures text][$\alpha-\alpha_{\text{exact}}$ convergence]{
\includegraphics[width=0.4\textwidth]{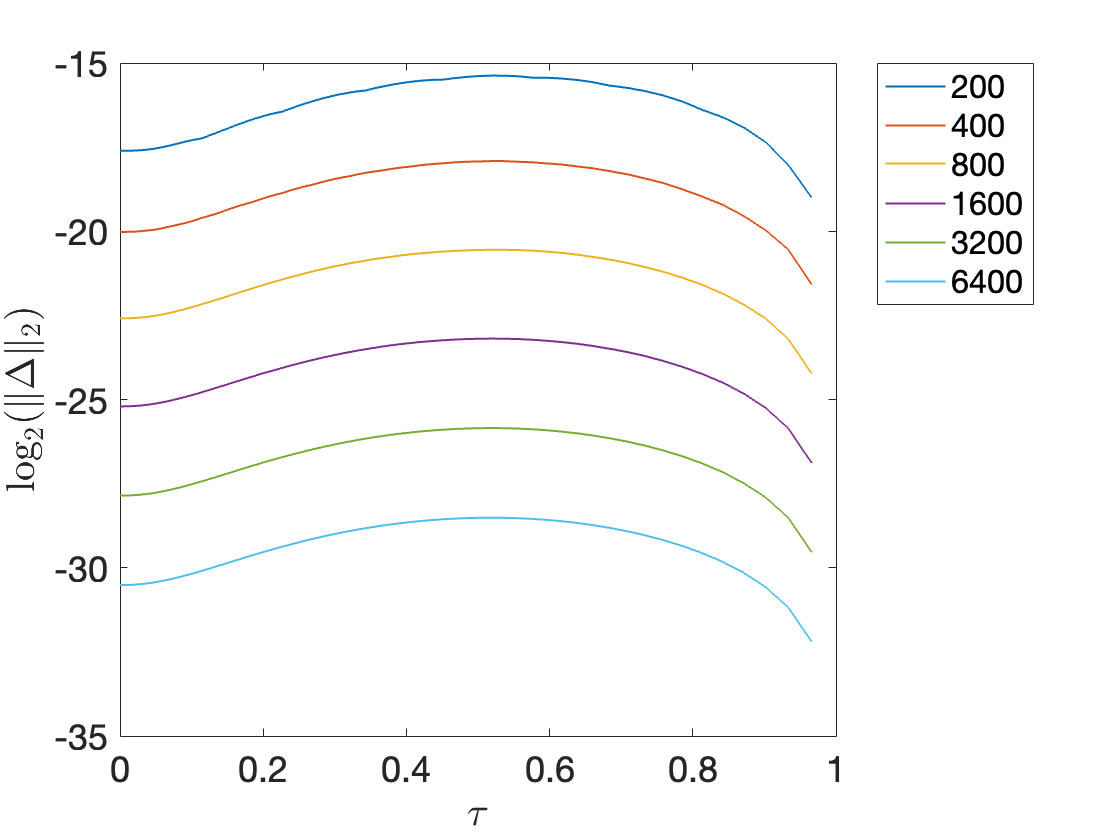}
\label{fig:subfigAlphaexact}}
\subfigure[Subfigure 2 list of figures text][$\rhot - \rhot_{\text{exact}}$ convergence]{
\includegraphics[width=0.4\textwidth]{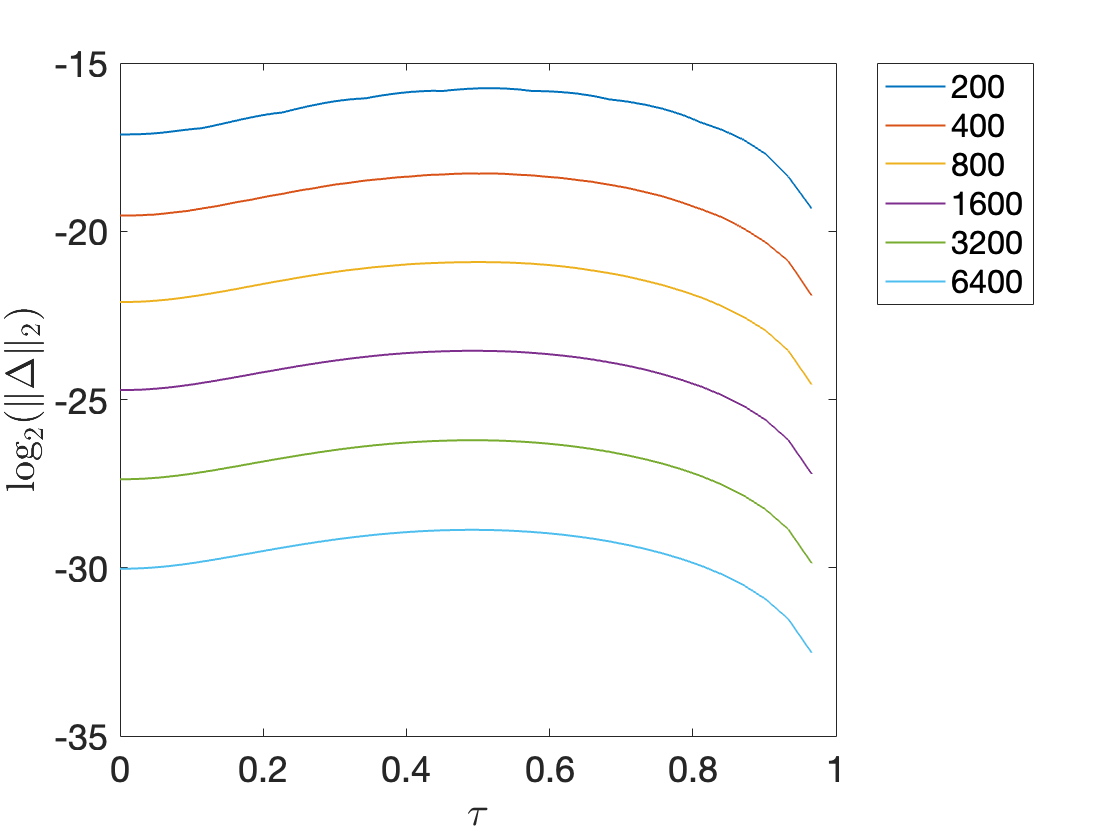}
\label{fig:subfigPexact}}
\caption{Convergence plots of the $L_{2}$ norm of $\alpha-\alpha_{\text{exact}}$ and $\rhot - \rhot_{\text{exact}}$, $K=0.4$, $\Lambda = 1$. The system was evolved until $\tau=0.002$.}
\end{figure}

A simple way to test the validity of our code is to compare our numerical solution with the FLRW solution \eqref{eqn:solnhomog}. For this convergence test, we employ the following initial data 
\begin{equation*}
\begin{aligned}
\rhot &= 1, \\
\alpha &= \frac{1}{2}\log\left(\frac{3}{\Lambda+1}\right), \\ 
A&=A_{1}=A_{0}=U=U_{1}=U_{0}=\eta=\tilde{v}_{1}=0.
\end{aligned}
\end{equation*}
Once again, we observe the expected second order convergence, shown for $\alpha$ and $\rhot$ in Figures \ref{fig:subfigAlphaexact} and \ref{fig:subfigPexact}, respectively.
\FloatBarrier
\subsection{Numerical Behaviour}
\label{sec:numericalbehaviour}
\begin{figure}[htbp]
\centering
\subfigure[Subfigure 1 list of figures text][$\tau = 0.001$]{
\includegraphics[width=0.3\textwidth]{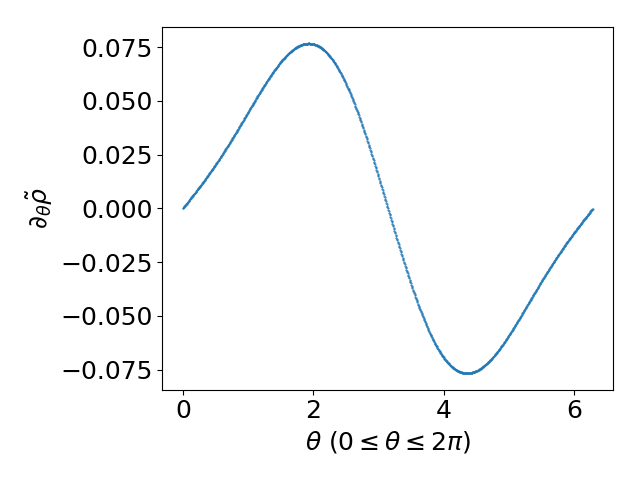}
\label{fig:Dxrho_t100}}
\subfigure[Subfigure 2 list of figures text][$\tau = 5.55 \times 10^{-6}$]{
\includegraphics[width=0.3\textwidth]{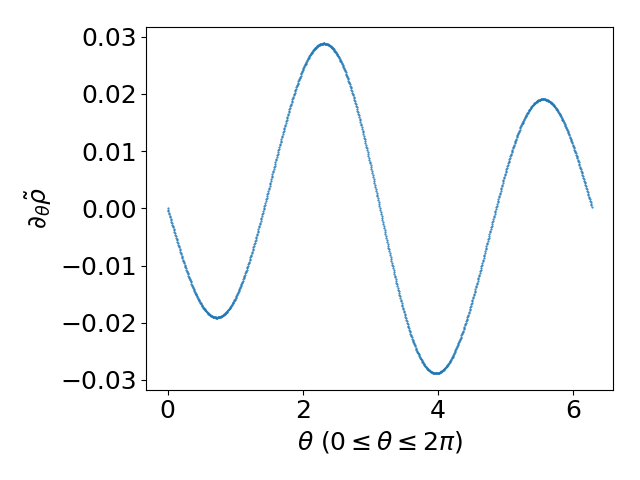}
\label{fig:Dxrho_t200}}
\subfigure[Subfigure 2 list of figures text][$\tau = 3.1 \times 10^{-8}$]{
\includegraphics[width=0.3\textwidth]{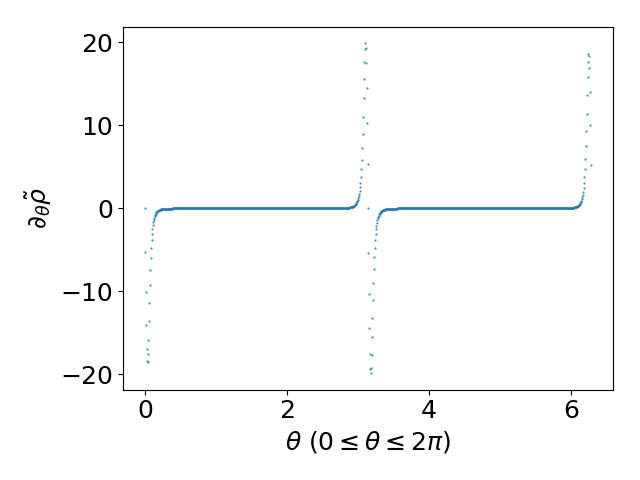}
\label{fig:Dxrho_t350}}
\caption{Plots of the derivative of the re-scaled density, $\del_{\theta}\rhot$, at various times. $N=1000$, $K=0.5$, $\Lambda = 1$.}
\label{fig:Dxrhoplots}
\end{figure}

We now examine the behaviour of numerical solutions of \eqref{eqn:EinsteinEuler1}-\eqref{eqn:EinsteinEuler2} with initial data of the form \eqref{eqn:numericalID}. From our numerical simulations, we observe that the asymptotic behaviour of the fluid variables and the fractional density gradient are broadly consistent with what was observed in \cite[\S 3.2]{Marshalloliynyk:2022} in the fixed background spacetime case. More specifically, for the full parameter range $1/3<K<1$ and all choices of the initial data with $a,$ $b,$ $c$, and $d$ sufficiently small, we observe that all the gravitational and fluid variables, with the exception of $\rhot$, remain bounded. It is unclear from our numerical solutions whether $\rhot$ remains bounded at timelike infinity. On the other hand, the spatial derivative of the density, $\del_{\theta}\rhot$, always develops steep gradients at finitely many points and becomes unbounded as $\tau\searrow 0$ for all $K\in(1/3,1)$, shown in Figure \ref{fig:Dxrhoplots}, indicating that the system is unstable.

In turn, this means the fractional density gradient, which is a measure of deviation from spatial homogeneity, also forms steep gradients and becomes unbounded as $\tau\searrow 0$, where we note $\tau=0$ corresponds to future timelike infinity. We present plots of the fractional density gradient in Section~\ref{sec:density_contrast}.

\subsubsection{Asymptotic Behaviour and Approximations}\label{sec:asympbehav}
\begin{figure}[htbp]
\centering
\subfigure[Subfigure 1 list of figures text][$\tau = 0.001$]{
\includegraphics[width=0.3\textwidth]{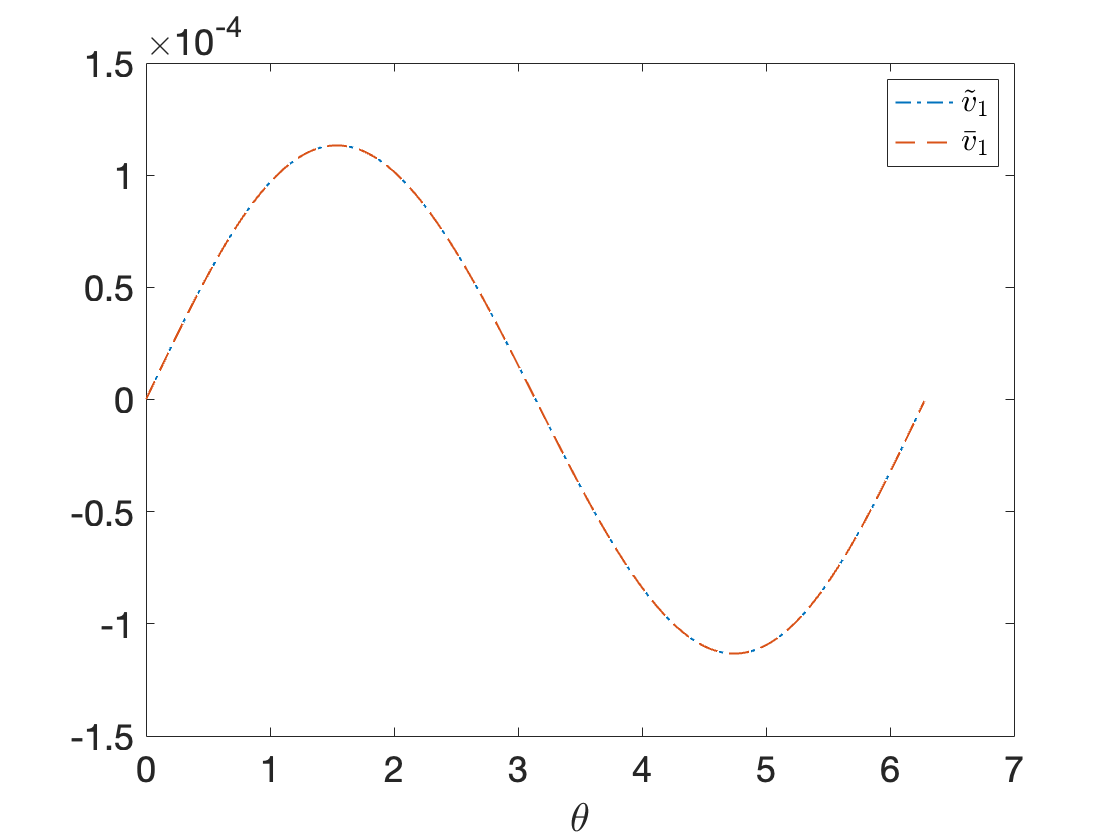}
\label{fig:subfigPDE155}}
\subfigure[Subfigure 2 list of figures text][$\tau = 5.55 \times 10^{-6}$]{
\includegraphics[width=0.3\textwidth]{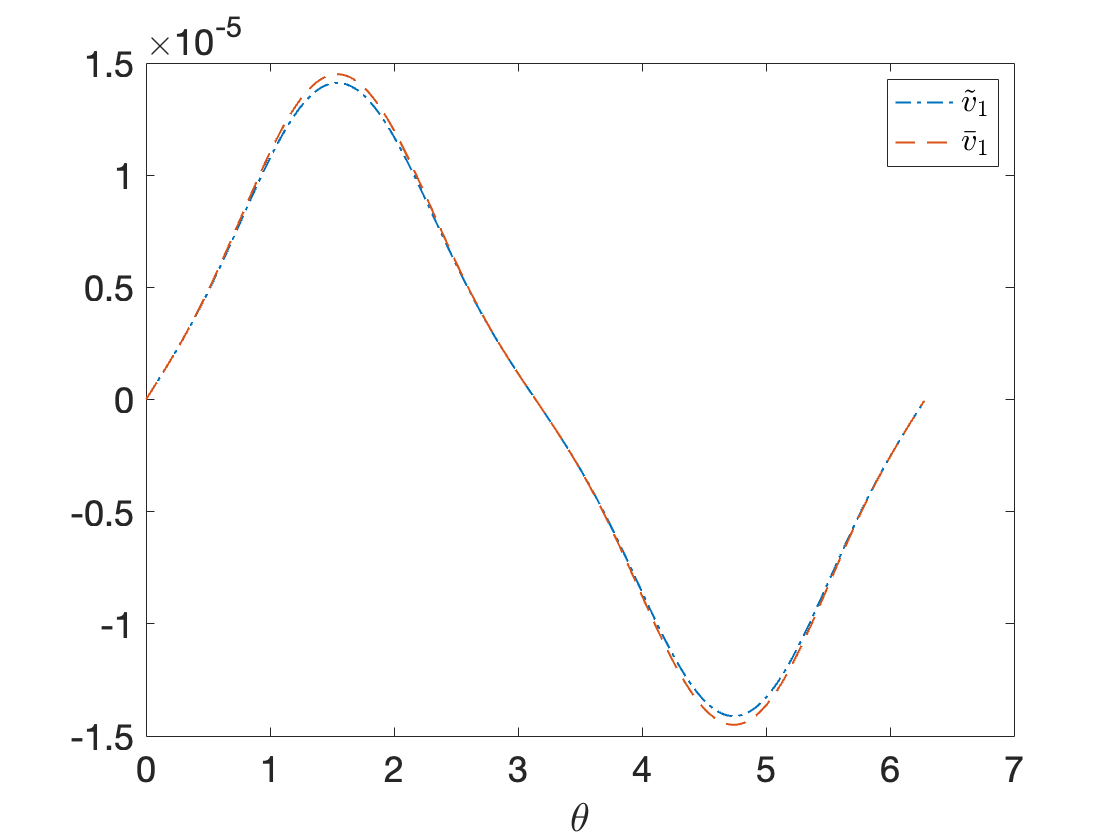}
\label{fig:subfigPDE200}}
\subfigure[Subfigure 2 list of figures text][$\tau = 9.79 \times 10^{-10}$]{
\includegraphics[width=0.3\textwidth]{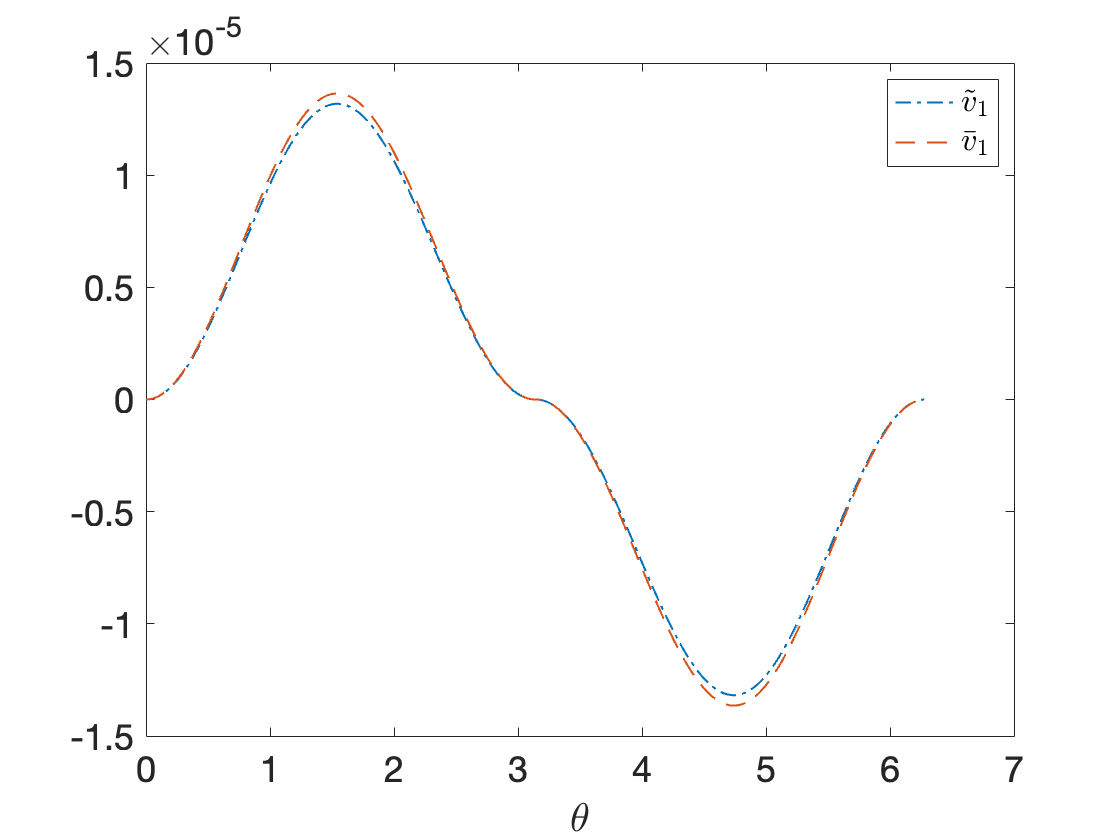}
\label{fig:subfigPDE900}}
\caption{Comparison of full Einstein-Euler solution $\tilde{v}_{1}$ (in blue) and asymptotic solution $\bar{v}_{1}$ (in orange) at various times. $\tilde{\tau}_{0} = 0.001$, $N=1000$, $K=0.5$, $\Lambda = 1$.}
\label{fig:asymptoticcompare1}
\end{figure}

In \cite[\S 3.2]{Marshalloliynyk:2022}, it was observed that the fluid variables displayed ODE-like behaviour at late times. This can also be seen for the metric and fluid variables in our simulations. In particular, we observe that, near $\tau=0$, solutions to the Gowdy-Euler equations are remarkably well approximated by solutions to the asymptotic system
\begin{align}
\label{eqn:A0homog}
\del_{\tau}A_{0} =&\; \alpha_{0}A_{0}+\frac{1}{\tau}(4\tau A_{1}U_{1}+2A_{0}-4\tau A_{0}U_{0}), \\
\del_{\tau}A_{1} =&\; \alpha_{0}A_{1}, \\
\del_{\tau}U_{0} =&\; \alpha_{0}U_{0} -\frac{1}{2\tau}(e^{4U}\tau A_{1}^{2}-e^{4U}\tau A_{0}^{2}-4U_{0}),  \\
\del_{\tau}U_{1} =&\; \alpha_{0}U_{1}, \\
\del_{\tau}\rhot =&\; \frac{1}{-e^{2U}(-1+K)\tilde{v}_{1}^{2}+e^{2\eta}\tau^{2\mu}}\Bigg(\tau^{-1}(1+K)\rhot\bigg(e^{2U}(-1+3K)\tilde{v}_{1}^{2} \nonumber \\
&+e^{2\eta}\tau^{1+2\mu}(U_{0}-\del_{\tau}\eta)+\frac{e^{2\eta}\tilde{v}_{1}\tau^{\mu+1} U_{1}}{\sqrt{e^{2(\eta-U)}+\tilde{v}_{1}^{2}\tau^{-2\mu}}}\bigg)\Bigg),  \\
\del_{\tau}\tilde{v}_{1} =&\; \frac{1}{\tau \Big(\tau^{2 \mu } e^{2 \eta}-(K-1) \tilde{v}_{1}^2 e^{2 U}\Big)}\Bigg(e^{-\alpha } \bigg(\tilde{v}_{1} \tau^{2 \mu } e^{\alpha +2 \eta} \Big(K \tau (\del_{\tau}\eta-U_{0})-3K+\mu +1\Big) \nonumber \\
& -\tilde{v}_{1}^3 \Big(K (\mu +3)-\mu -1\Big) e^{\alpha +2 U}-U_{1} \tau^{3 \mu +1} e^{2 \eta+\alpha} \sqrt{(e^{2 \eta-2 U}+\tilde{v}_{1}^2 \tau^{-2 \mu })}\bigg)\Bigg), \\ 
\del_{\tau}\alpha =&\;\frac{\left(2 \Lambda -(K-1) \rhot \tau^{3 K+3}\right) e^{ 2 (\alpha -U+\eta)}-6}{2 \tau},  \\
\del_{\tau}\eta=&\; \frac{1}{8}e^{-2U}\tau^{-1-2\mu}\Bigg(-4e^{2(U+\alpha)}(1+K)\rhot \tilde{v}_{1}^{2}\tau^{3+3K}-\tau^{2\mu}\bigg(4\left(-3e^{2U}+e^{2(\eta+\alpha)}(\Lambda+\rhot \tau^{3+3K})\right) \nonumber \\
&+e^{2U}\tau\Big(e^{4U}\tau(A_{1}^{2}+A_{0}^{2})-8\del_{\tau}U+4\tau\big(U_{1}^{2}+U_{0}^{2}\big)\Big)\bigg)\Bigg),  \\
\del_{\tau}A =&\;A_{0}, \\ 
\label{eqn:Uhomog}
\del_{\tau}U =&\;U_{0}.  
\end{align}
This system is obtained by setting the spatial derivative terms in  \eqref{eqn:EinsteinEuler1}-\eqref{eqn:EinsteinEuler2} to zero. 
We have tested the agreement between solutions of the full Einstein-Euler equations \eqref{eqn:EinsteinEuler1}-\eqref{eqn:EinsteinEuler2} and the asymptotic system \eqref{eqn:A0homog}-\eqref{eqn:Uhomog} using the following procedure, which is similar to the one employed in \cite[\S 3.2.2]{Marshalloliynyk:2022}:
\begin{enumerate}[(i)]
    \item Generate a numerical solution $(A,A_{1},A_{0},U,U_{1},U_{0},\alpha,\eta,\rhot,\tilde{v}_{1})$ of the Einstein-Euler equations \eqref{eqn:EinsteinEuler1}-\eqref{eqn:EinsteinEuler2} from initial data specified at $\tau_{0}>0$.
    \item Fix a time\footnote{It is worth noting that  the value of $\tilde{\tau}_{0}$ increases as $K \nearrow 1$.} $\tilde{\tau}_{0}$ when the solution from step (i) appears to be first dominated by ODE behaviour. 
    \item Fix initial data for the asymptotic system \eqref{eqn:A0homog}-\eqref{eqn:Uhomog} at $\tau = \tilde{\tau}_{0}$ by setting 
    \begin{align*}  (\bar{A}_{I},\bar{A}_{1,I},\bar{A}_{0,I},\bar{U}_{I},\bar{U}_{1,I},\bar{U}_{0,I},\bar{\alpha}_{I},\bar{\eta}_{I},\rhob_{I},\bar{v}_{1,I})= (A,A_{1},A_{0},U,U_{1},U_{0},\alpha,\eta,\rhot,\tilde{v}_{1})|_{\tilde{\tau}_{0}}.
    \end{align*}
    \item Numerically solve the asymptotic system \eqref{eqn:A0homog}-\eqref{eqn:Uhomog} using the initial data from (iii) to obtain a solution $(\bar{A},\bar{A}_{1},\bar{A}_{0},\bar{U},\bar{U}_{1},\bar{U}_{0},\bar{\alpha},\bar{\eta},\rhob,\bar{v}_{1})$.
    \item Compare the solutions $(\bar{A},\bar{A}_{1},\bar{A}_{0},\bar{U},\bar{U}_{1},\bar{U}_{0},\bar{\alpha},\bar{\eta},\rhob,\bar{v}_{1})$ and $(A,A_{1},A_{0},U,U_{1},U_{0},\alpha,\eta,\rhot,\tilde{v}_{1})$ on the region $(0,\tilde{\tau}_{0})$.
\end{enumerate}
Following this process, we observe that the metric variables $A,U,A_{1},U_{1},\alpha,$ and $\eta$ become effectively constant for $\tau \in (0,\tilde{\tau}_{0})$ and are indistinguishable from the corresponding asymptotic solution $\bar{A},\bar{U},\bar{A}_{1},\bar{U}_{1},\bar{\alpha},$ and $\bar{\eta}$, while the variables $A_{0}$, $U_{0}$, $\bar{A}_{0}$, and $\bar{U}_{0}$ all rapidly decay to zero. On the other hand, the fluid variables, $\tilde{v}_{1}$ and $\rhot$, display significantly more dynamic, but still ODE-dominated,  behaviour before converging to fixed functions for $\tau\in (0,\tilde{\tau}_{0})$. In particular, the fluid variables closely match their asymptotic counterparts, shown for $\tilde{v}_{1}$ in Figure \ref{fig:asymptoticcompare1}. Furthermore, we note that $\tilde{v}_{1}$ and $\bar{v}_{1}$ show strong agreement even at the locations where spike points form in the fractional density gradient, cf.\ Figure~\ref{fig:DCPlots}.

\subsubsection{Behaviour of the fractional density gradient} 
\label{sec:density_contrast}

\begin{figure}[htbp]
\centering
\subfigure[Subfigure 1 list of figures text][$\tau = 0.001$]{
\includegraphics[width=0.3\textwidth]{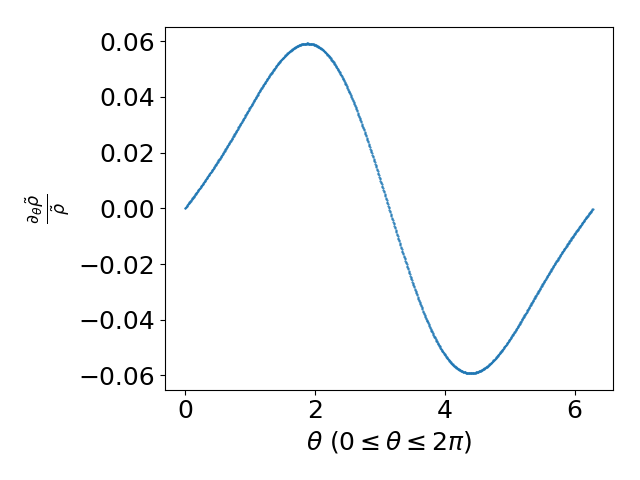}
\label{fig:subfigDC_t200}}
\subfigure[Subfigure 2 list of figures text][$\tau = 5.55 \times 10^{-6}$]{
\includegraphics[width=0.3\textwidth]{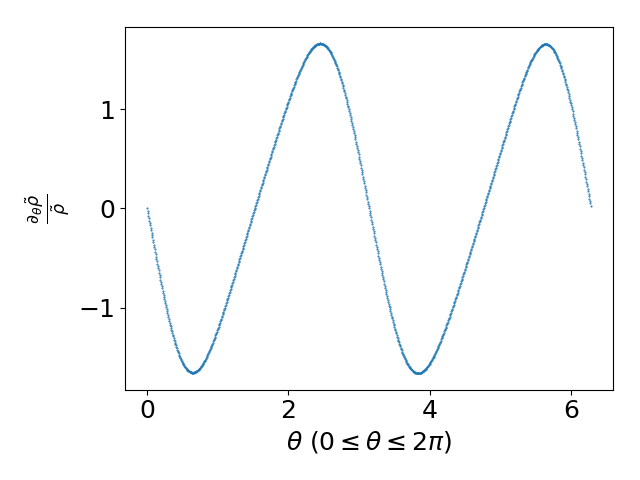}
\label{fig:subfigDC_t350}}
\subfigure[Subfigure 2 list of figures text][$\tau = 3 \times 10^{-8}$]{
\includegraphics[width=0.3\textwidth]{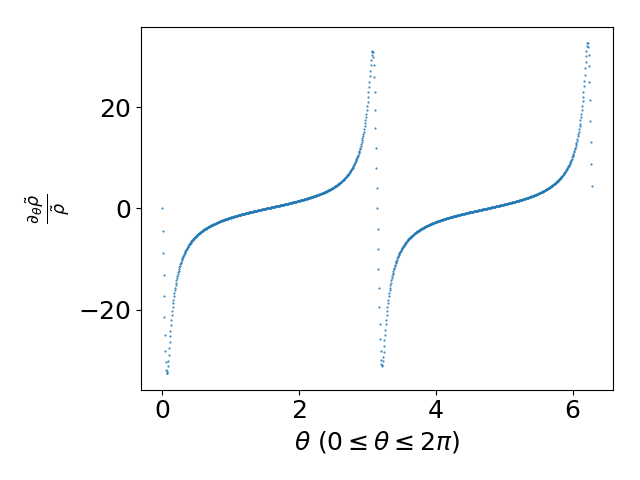}
\label{fig:subfigDC_t500}}
\caption{fractional density gradient $\frac{\del_{\theta}\rho}{\rho}$ at various times. $N=1000$, $K=0.5$, $\Lambda = 1$.}
\label{fig:DCPlots}
\end{figure}

The fractional density gradient is, by definition, $\frac{\del_{\theta}\rho}{\rho}$. In terms of the re-scaled density \eqref{eqn:rhotdef}, it is given by
\begin{align*}
    \frac{\del_{\theta}\rho}{\rho} = \frac{\del_{\theta}\rhot}{\rhot}.
\end{align*}
Using this relation, we observe from the numerical simulations that the fractional density gradient develops steep gradients and blows-up at $\tau=0$ at isolated spatial points, shown in Figure \ref{fig:DCPlots}. 
As discussed in the introduction, this singular behaviour was anticipated by Rendall in \cite{Rendall:2004}. 
The fractional density gradient blow-up at timelike infinity also indicates an instability in the sense that the fractional density gradient of the perturbed solutions does not remain uniformly bounded no matter how close the initial data is chosen to FLRW initial data. It should be noted, however, that the blow-up of the fractional density gradient is more apparent as the size of $K$ increases. In particular, for values of $K$ close to $1/3$ one needs to choose initial data with larger values of $a$, $b$, $c$, and $d$ to observe the blow-up within the timespan of our numerical evolutions. 

\begin{figure}[htbp]
\centering
\subfigure[Subfigure 1 list of figures text][$\tau = 0.001$]{
\includegraphics[width=0.3\textwidth]{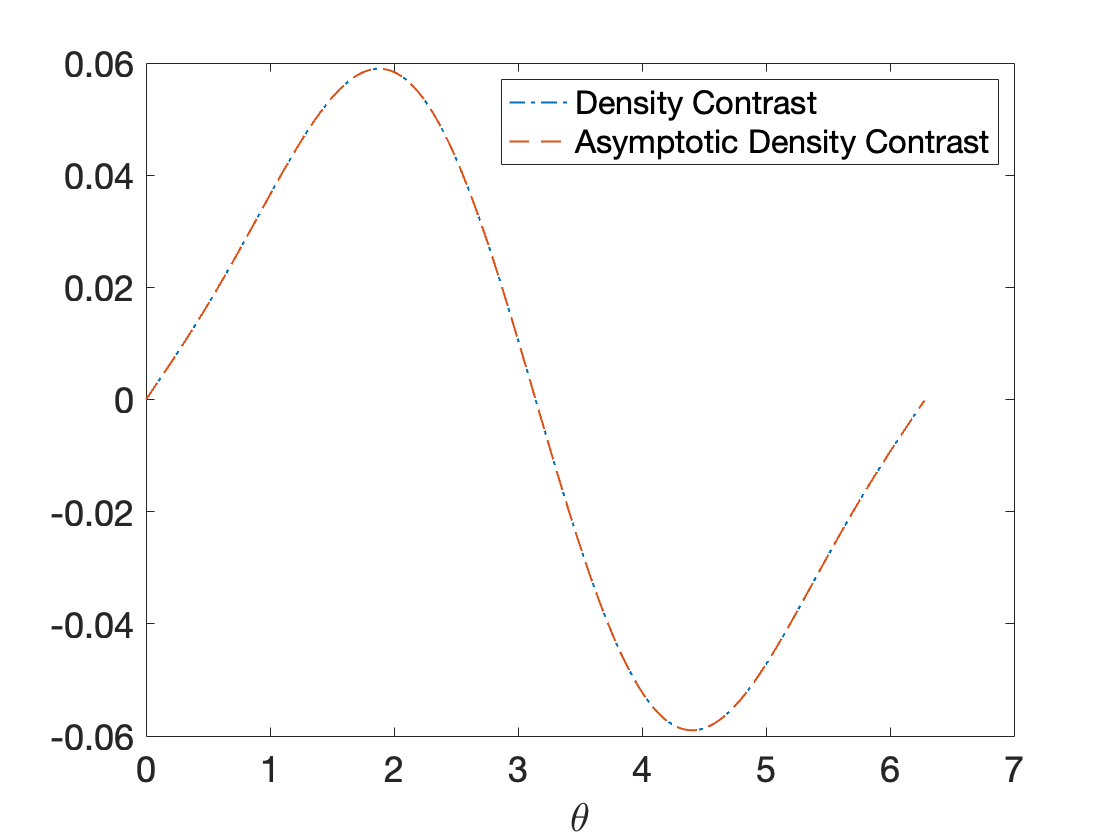}
\label{fig:subfigDC_compare155}}
\subfigure[Subfigure 2 list of figures text][$\tau = 5.55 \times 10^{-6}$]{
\includegraphics[width=0.3\textwidth]{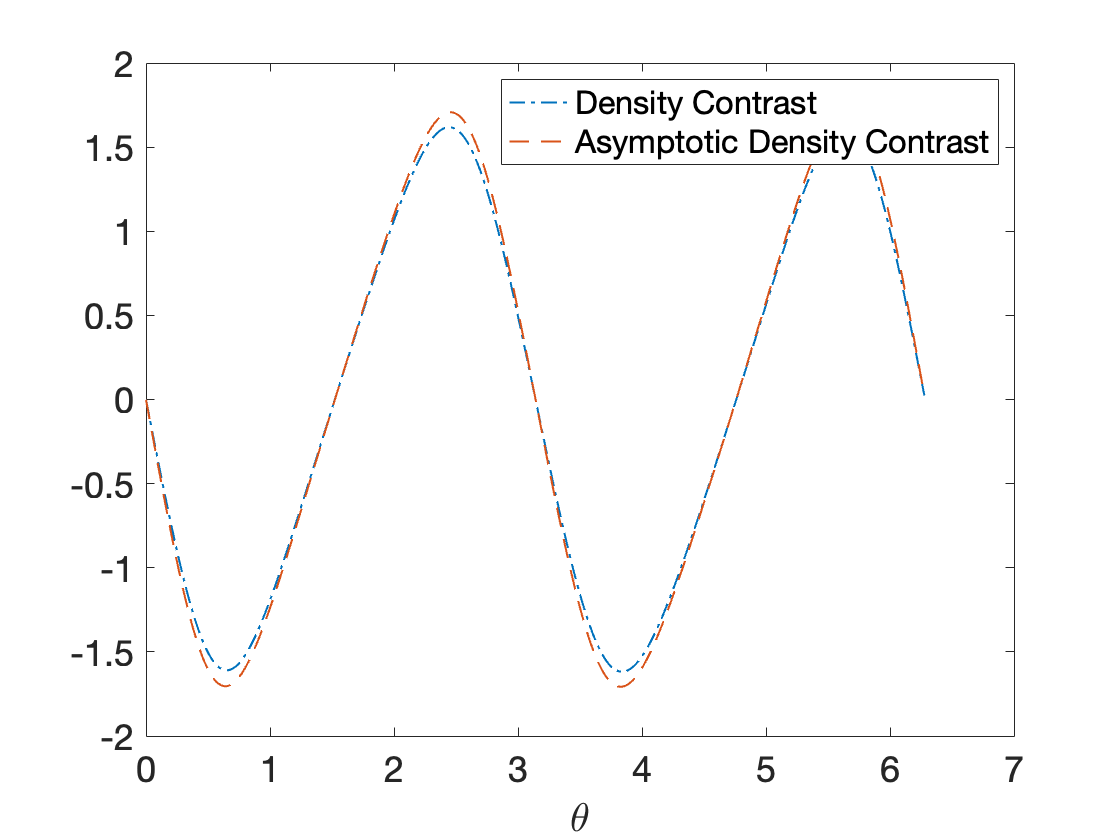}
\label{fig:subfigDC_compare250}}
\subfigure[Subfigure 2 list of figures text][$\tau = 9.79 \times 10^{-10}$]{
\includegraphics[width=0.3\textwidth]{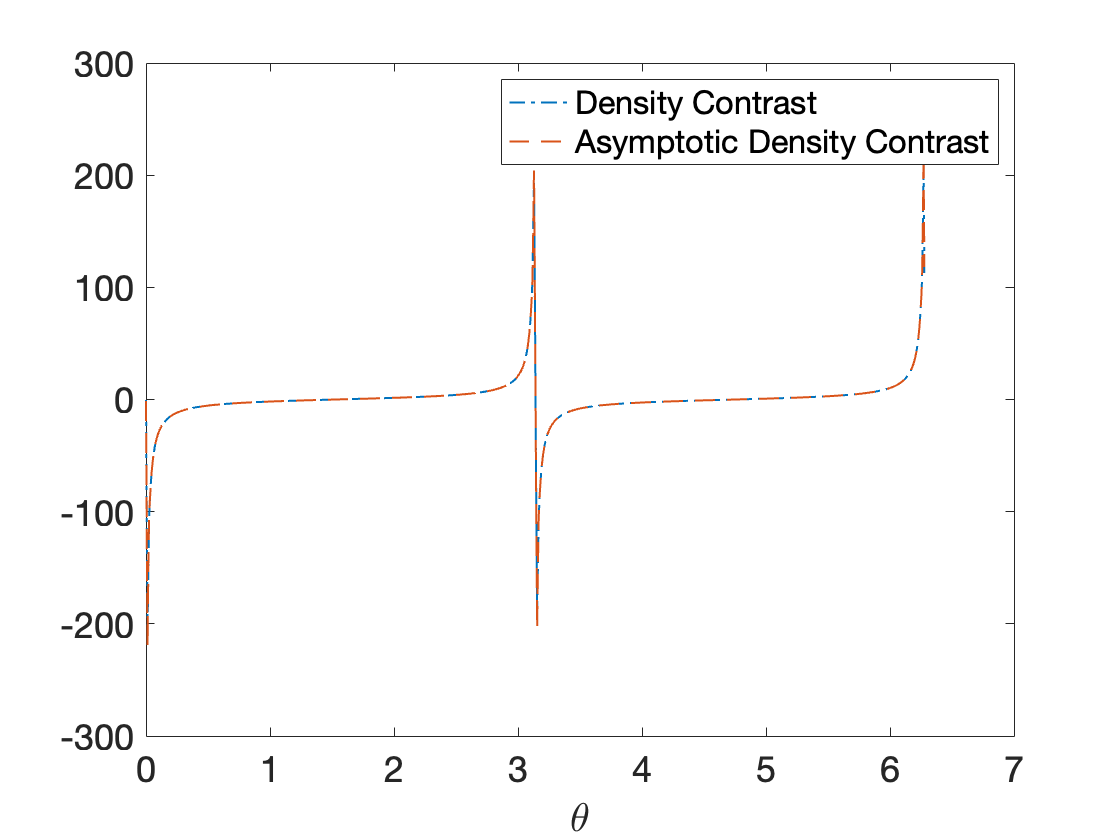}
\label{fig:subfigDC_compare350}}
\caption{Comparison between the full Einstein-Euler (in blue) and asymptotic fractional density gradient (in orange) at various times. $\tilde{\tau}_{0}=0.001, N=1000$, $K=0.5$, $\Lambda = 1$.}
\label{fig:AsymptoticDCPlots}
\end{figure}

Finally, as in Section \ref{sec:asympbehav}, we can compare the fractional density gradient computed from solutions of the full Einstein-Euler equations \eqref{eqn:EinsteinEuler1}-\eqref{eqn:EinsteinEuler2} with the fractional density gradient generated from the asymptotic system. Once again the full Einstein-Euler and asymptotic plots are almost indistinguishable, shown in Figure \ref{fig:AsymptoticDCPlots}.  

\FloatBarrier

\section{Discussion}
The aim of this work was to numerically study nonlinear perturbations of FLRW solutions to the Einstein-Euler equations under a Gowdy symmetry assumption and linear equation of state $p=K\rho$ for $K\in(1/3,1)$. In particular, our objective was to determine whether the blow-up of the fractional density gradient $\frac{\del_{\theta}\rho}{\rho}$ at isolated spatial points at timelike infinity, anticipated by Rendall~\cite{Rendall:2004} and subsequently numerically observed for the relativistic Euler equations on an exponentially expanding FLRW spacetime in \cite{Marshalloliynyk:2022}, also occurs when coupling to Einstein gravity is included. We have numerically solved the Einstein-Euler equations using a standard second-order Runge-Kutta method in time and second-order central finite differences to discretise spatial derivatives. The expected second order accuracy of this implementation was confirmed by our convergence tests. Using this numerical scheme, we found that the fractional density gradient blows up at finitely many spatial points as $\tau \searrow 0$ for all $K\in(1/3,1)$ and all choices of initial data that are sufficiently close to FLRW initial data and for which $\tilde{v}_{1}$ crosses zero somewhere on the initial hypersurface. These results are consistent with the fractional density gradient blow-up scenario put forth by Rendall. 

In the influential article \cite{Starobinskii:1983}, asymptotic expansions near future timelike infinity for vacuum and perfect fluid cosmologies with a positive cosmological model were derived; see also \cite{lim_et_al:2004,Rendall:2004} for later work. These expansions indicate that one should expect that solutions to cosmological models with a positive cosmological constant will asymptotically isotropize and approach de Sitter spacetime in a suitable sense, which is consistent with the cosmic no-hair conjecture \cite{HawkingMoss:1982}. This expectation was later strengthened by the proof of asymptotic isotropization and convergence to de Sitter spacetime for spatially homogeneous cosmological models by Wald \cite{Wald:1983} and the
generalization of this result to inhomogeneous cosmological models under a negative spatial scalar curvature assumption \cite{JensenStein-Schabes:1987}. Furthermore, the rigorous stability results established in the articles \cite{andreasson2016,Fournodavlos:2022,Friedrich:1986,Friedrich:1991,Friedrich:2017,HadzicSpeck:2015,LubbeKroon:2013,Oliynyk:CMP_2016,ringstrom2015b,RodnianskiSpeck:2013,Speck:2012} provide further compelling support for this viewpoint. Together, all of these results have led to an expected late time behavior for cosmologies with a positive cosmological constant that involves asymptotic isotropization and convergence to de Sitter spacetime.  For an extended discussion on the status of the asymptotic isotropization of cosmologies, see the article \cite{lim_et_al:2004}.

Now, as first observed by Rendall \cite{Rendall:2004}, the vanishing of the (rescaled) spatial fluid velocity
at any point at timelike infinity is an obstruction to existence of asymptotic expansions of the type derived in \cite{Starobinskii:1983}. Moreover, he conjectured that the vanishing of the (rescaled) spatial fluid velocity at some point at timelike infinity would lead to
blow-up of the fractional density gradient at timelike infinity for $1/3<K<1$. This is exactly what we observe in our numerical simulations, namely, that the rescaled spatial fluid velocity, which is $\vt_1$ in our notation, vanishes at $\tau=0$ at a finite set of spatial points and that the fractional density gradient $\frac{\del_{\theta}\rho}{\rho}$ 
develops steep gradients near those same spatial points and becomes unbounded there as $\tau \searrow 0$.
Thus, it is in this sense that the cosmological solutions studied in this article do not display the expected behaviour and are of possible physical interest.

We have also observed that, for initial data suitably close to spatially homogeneous initial data, solutions display ODE-like behaviour at late times analogous to the behaviour found in \cite{Marshalloliynyk:2022}. For cosmological solutions that admit asymptotic expansions of the type considered in \cite{Starobinskii:1983}, it is not difficult to verify that they are dominated by ODE behavior near future timelike infinity. We also note ODE dominated behaviour near timelike infinity for
nonlinear perturbation of FLRW solutions to the Einstein-Euler equations with a positive cosmological constant can be rigorously established using the stability results from the articles \cite{Friedrich:2017,HadzicSpeck:2015,LubbeKroon:2013,Oliynyk:CMP_2016,RodnianskiSpeck:2013,Speck:2012}. The ODE dominance for these types of solutions is a consequence of the asymptotic isotropization of the solutions, which means that the spatial derivative terms in the evolution equations make a negligible contribution to solutions near timelike infinity. Thus an approximation, which gets better the closer to future timelike infinity where the initial data is prescribed, can be generated
by solving the system of ODEs obtained by omitting the spatial derivative terms from the evolution equations.   

What is surprising is that the ODE dominance continues to be true for the solutions considered in this article. The reason that this is surprising is that the solutions are highly non-homogeneous near spatial infinity and so one would expect that the derivative terms in the evolution equations should introduce non-negligible effects. While this is the case near a finite number of points, it is remarkable that the ODE approximation remains valid everywhere else, in particular, even in regions very close to these exceptional points. It is conceivable that the dynamics near those exceptional points resembles that of \emph{spikes} that have been identified as a common feature near big bang singularities \cite{berger1998c,berger1993,coley2015,heinzle2012a,lim2008,lim2009,rendall2001}. We plan on investigating this possible connection in future work.

There are several directions for future research to take. An obvious next step would be to remove the Gowdy symmetry assumption and study the full 3+1 system. Additionally, while we believe the initial data we have studied is reasonably `generic', it would be interesting to test a wider variety of initial conditions to see what, if any, impact this has on the behaviour of solutions.

\bibliographystyle{amsplain}
\bibliography{refs}

\end{document}